\documentclass[%
 notitlepage,
 twocolumn,
superscriptaddress,
showpacs,
amsmath,amssymb,
aps,
pra,
floatfix,
longbibliography
]{revtex4-2}

\usepackage[T1]{fontenc}
\usepackage{float}
\usepackage{amsmath}
\usepackage{amssymb}
\usepackage{graphicx}
\usepackage{esint}
\PassOptionsToPackage{normalem}{ulem}
\usepackage{ulem}
\usepackage{xcolor}
\usepackage{hyperref}
\hypersetup{
	citecolor=blue,
	colorlinks=true,
	urlcolor=magenta,
}

\makeatletter

\newcommand{\VILNIUS}{Institute of Theoretical Physics and Astronomy, Faculty of Physics, Vilnius University, Saul\.etekio 3, LT-10257, Vilnius, Lithuania}

\makeatother

\begin{document}
\title{Two dimensional sub-wavelength topological dark state lattices}
% \author{Domantas Burba and Gediminas Juzeli\={u}nas}

%
\author{D. Burba}
\affiliation{\VILNIUS}
\author{G. Juzeli\=unas}
\affiliation{\VILNIUS}

% \lyxaddress{Institute of Theoretical Physics and Astronomy, Department of Physics,
% Vilnius University, Saul\.{e}tekio 3, LT-10257 Vilnius, Lithuania}

% \db{Needs to be shortened.} \gj{I think the length of the abstract is OK. }
\begin{abstract}
We present a general framework
for engineering two-dimensional sub-wavelength topological optical
lattices using spatially dependent atomic dark states in a $\Lambda$-type
configuration of the atom-light coupling. By properly designing the
spatial profiles of the laser fields inducing coupling between the
atomic internal states, we show how to generate sub-wavelength Kronig-Penney-like
geometric scalar potential accompanied by narrow and strong patches
of the synthetic magnetic field localized in the same areas as the
scalar potential. These sharply peaked magnetic fluxes are compensated
by a smooth background magnetic field of opposite sign, resulting
in zero net flux per unit cell while still enabling topologically
nontrivial band structures. Specifically, for sufficiently narrow
peaks, their influence is minimum, and the behavior of the system
in a  remaining smooth background magnetic field resembles the Landau problem,
allowing for the formation of nearly flat energy bands with unit Chern
numbers. Numerical analysis confirms the existence of ideal Chern
bands and the robustness of the topological phases against non-adiabatic
effects and losses. This makes the scheme well-suited for simulating
quantum Hall systems and fractional Chern insulators in ultracold
atomic gases, offering a platform for exploring strongly correlated
topological phases with high  tunability.
\end{abstract}

\maketitle

\section{Introduction}

% \textbf{arxiv citations don't have number, change revtex}
% \db{Figures need to be vectorized and letters in figures increased.}
% \gj{References should appear in a consecutive order, not like it is now.}

\textcolor{black}{Topological effects play an important roles in various areas of physics, including 
%\it{inter alia}
condensed matter~\cite{QiZhangRMP2011, HasanKaneRMP2010}, ultracold atoms~\cite{Dalibard11RMP,Goldman2014,Aidelsburger2018Physique,Galitski19PT,Cooper2019RMP}, optics and photonics~\cite{Ozawa2019RMP,Shen2024,Sonja2023}, acoustics~\cite{Xue2022NatRev}
and its non-linear realizations ~\cite{XinxinGuo2025}.}
Ultracold atoms represent a flexible platform for simulating topological
and many-body phenomena of condensed matter and high-energy physics
\cite{Lewenstein2007,Bloch2008,Dutta15RoPP,Gross-Bloch17Science,Takahashi20NatRev}.
The use of atomic dark states offers additional possibilities for such simulations.
The dark states are long-lived superpositions of atomic internal ground
states immune to atom-light coupling \cite{Scully2008,Bergman2017}.
Making the dark states position-dependent, one can create unconventional
optical lattices or generate a synthetic magnetic field for ultracold
atoms adiabatically following the dark states \cite{Dalibard11RMP,Goldman2014}.
In particular, using this method, one can create a one-dimensional
(1D) array of sub-wavelength barriers \cite{Zoller2016,Jendrzejewski2016,Zubairy2020,Kubala2021,Gvozdiovas2021},
thus providing optical lattices that were realized experimentally
\cite{Wang2018,Tsui2020}. Recently, two-dimensional (2D) dark-state
lattices were also considered \textcolor{black}{\cite{Gvozdiovas23PRA,Dalibard24arXiv,Cooper2025arXiv}.
In particular,} it was demonstrated that in addition to the geometric scalar potential,
the dark state atoms can also be affected by a geometric vector potential
corresponding to a non-zero magnetic field.  The magnetic flux produced
in this way is staggered \cite{Gvozdiovas23PRA}, but can have features
of the non-staggered flux for a very specific set of parameters \cite{Dalibard24arXiv}. 

Here we provide a general description of 2D topological dark state
lattices elucidating an interplay with the sub-wavelength lattices.
In particular, we demonstrate that one can create a toplogical 2D
Kronig-Penney like lattice which contains periodically distributed
narrow peaks of strong scalar potential and strong non-staggered magnetic
field, the latter   located at the same positions as the peaks of
the scalar potential.   Away from these subwavelength patches of
the strong magnetic field, there is a smooth magnetic field of the
opposite sign, compensating for the former peaks, so the total magnetic
flux over an elementary cell is zero. Nevertheless, the system supports
topological phases due to the flux variation over a unit cell, involving
large areas of smooth magnetic field. In particular, as the width of the subwavelength peaks of the magnetic
fluxes becomes sufficiently small, their influence diminishes and the problem reduces
to the motion of the particle in a nearly uniform background field
akin to the Landau problem for a particle moving in a constant magnetic
field~\textcolor{black}{\cite{Landau1930, landau1977quantum}}. This work paves the way for experimental exploration of topological
phases in 2D  sub-wavelength dark-state optical lattices, offering possibilities
for simulating quantum Hall systems, fractional Chern insulators and
related strongly correlated phases. 

The paper is organised as follows. In the next Section we introduce
the $\Lambda$ scheme of the atom-light coupling and define the dark
states which are immune to the atom-light coupling. Section~\ref{sec:Adiabatic-motion-in}
provides the general description of adiabatic motion of the dark state
atoms affected by the geometric scalar and vector potential, as well
as the corresponding magnetic field. Section~\ref{sec:Atom-light-coupling}
considers specific light fields for the $\Lambda$ scheme providing
a 2D  sub-wavelength topological lattice for the atomic motion in the dark state manifold.
%The results are discussed in the concluding Sec.~\ref{sec:Discussion}.
 The concluding Sec.~\ref{sec:Discussion} summarizes the findings.

\section{Formulation\label{sec:Formulation}}

\subsection{Hamiltonian}

% [H]
\begin{figure}[tbh!]
\centering
\includegraphics[width=0.98\columnwidth]{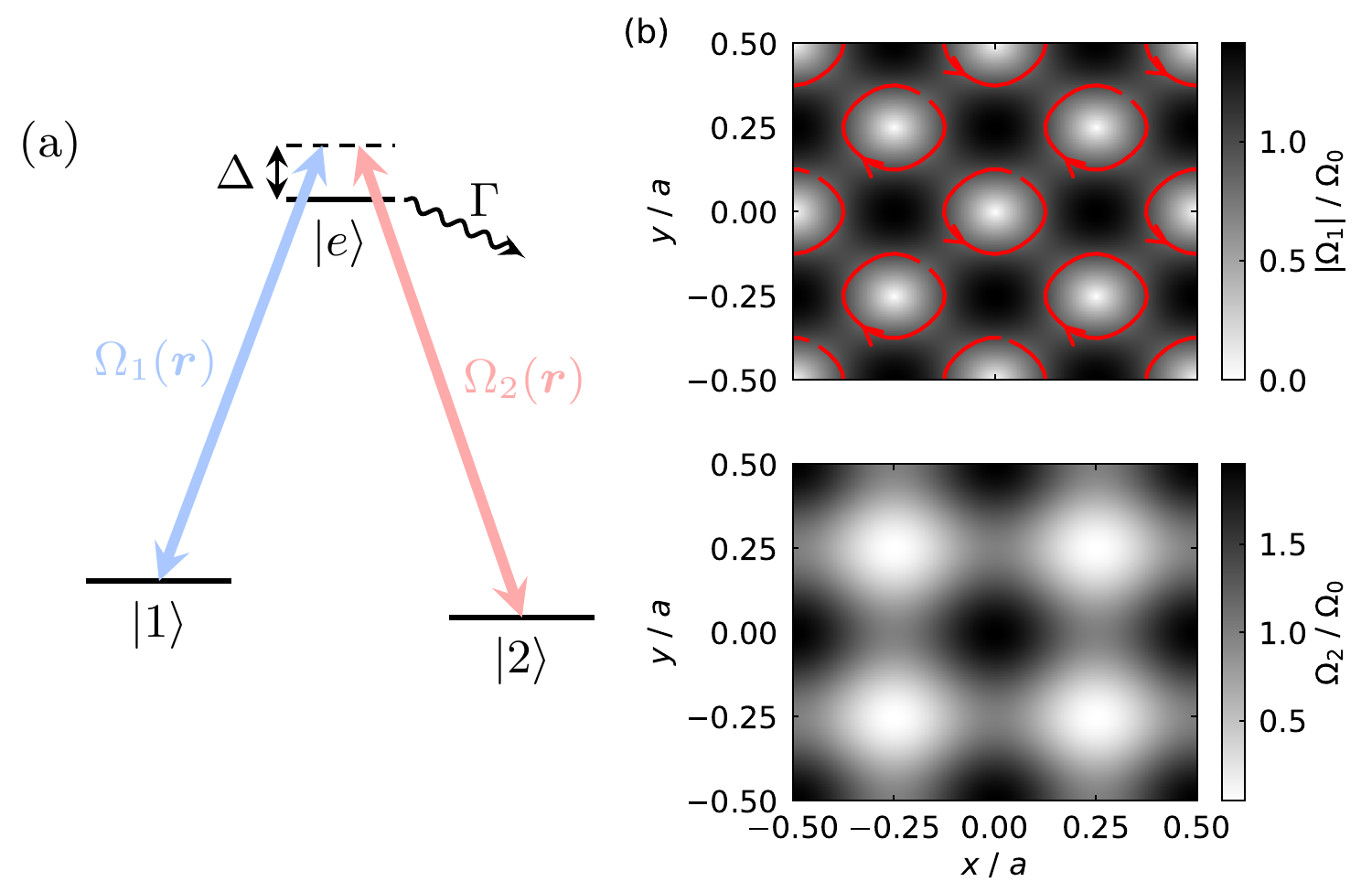}
\caption{
% \db{Looks too empty, need to add panel (b) with Omega1 and 2 dependence in space.}
(a) The Lambda scheme of the atom-light coupling. Two laser fields characterized
by Rabi frequencies $\Omega_{1}$ and $\Omega_{2}$ couple resonantly
(or nearly resonantly with a single photon detuning $\Delta$) two
atomic ground states $\left|1\right\rangle $ and $\left|2\right\rangle $
to a common excited state $\left|e\right\rangle $. (b) The position dependence of the  modulus of Rabi frequencies $\Omega_{1}$ and $\Omega_{2}$ given by by Eqs.~\eqref{eq:Omega_1-specific}-\eqref{eq:Omega_2-specific} for $\epsilon=1$ and $\nu=0.95$.  Arrows indicate the direction of the phase winding of  $\Omega_{1}$ around its zero points.}

\label{fig:Lambda}
\end{figure}

 Let us consider atoms in the $\Lambda$ configuration of energy levels
involving two ground states $\left|1\right\rangle $ and $\left|2\right\rangle $
and an excited state $\left|e\right\rangle $   depicted in Fig.~\ref{fig:Lambda}(a). Both ground states
are resonantly (or nearly resonantly with a single photon detuning
$\Delta$) coupled to the same excited state $\left|e\right\rangle $
by laser fields characterized by the Rabi frequencies $\Omega_{1}$
and $\Omega_{2}$.
%, as depicted in Fig.~\ref{fig:Lambda}.
The Rabi
frequencies $\Omega_{1}\equiv\Omega_{1}\left(\mathbf{r}\right)$
and $\Omega_{2}\equiv\Omega_{2}\left(\mathbf{r}\right)$ are generally
position-dependent,  as shown in Fig.~\ref{fig:Lambda}(b), but this $\mathbf{r}$-dependence will be
mostly kept implicit. Applying the rotating wave approximation \cite{Scully2008}, the Hamiltonian describing the atom-light coupling
is:

\begin{equation}
\hat{V}/\hbar=\left(-\Delta-\frac{i}{2}\Gamma\right)\left|e\right\rangle \left\langle e\right|+\sum_{j=1}^{2}\left(\frac{\Omega_{j}}{2}\left|e\right\rangle \left\langle j\right|+{\rm H.c.}\right)\,.\label{eq:V-definition}
\end{equation}
where an imaginary contribution $-i\Gamma/2$ is added to the energy
of the excited state to describe its decay at the rate $\Gamma$.

\subsection{Dark and bright state basis}

The atom-light Hamiltonian ~(\ref{eq:V-definition}) can be represented
in term of the excited and bright states 
\begin{equation}
\hat{V}/\hbar=\left(-\Delta-\frac{i}{2}\Gamma\right)\left|e\right\rangle \left\langle e\right|+\frac{\Omega}{2}\left(\left|e\right\rangle \left\langle \tilde{B}\right|+\left|\tilde{B}\right\rangle \left\langle e\right|\right)\,,\label{eq:V-definition-1}
\end{equation}
where the bright state $\left|\tilde{B}\right\rangle $ is a superposition
of the atomic ground states $\left|1\right\rangle $ and $\left|2\right\rangle $
directly coupled to the excited state $\left|e\right\rangle $: 
\begin{equation}
\left|\tilde{B}\right\rangle =\frac{1}{\Omega}\left(\Omega_{1}^{*}\left|1\right\rangle +\Omega_{2}^{*}\left|2\right\rangle \right)\,,\label{eq:B}
\end{equation}
with 
\begin{equation}
\Omega=\sqrt{\left|\Omega_{1}\right|^{2}+\left|\Omega_{2}\right|^{2}}\label{eq:Omega}
\end{equation}
being the total Rabi frequency. The third atomic state is the dark
state representing a superposition of ground states orthogonal to
the bright state 
\begin{equation}
\left|\tilde{D}\right\rangle =\frac{1}{\Omega}\left(\Omega_{2}\left|1\right\rangle -\Omega_{1}\left|2\right\rangle \right)\,.\label{eq:D}
\end{equation}
Both dark and bright states are position-dependent atomic dressed
states. The spatial variation of these states comes from the position-dependence
of the Rabi frequencies $\Omega_{1}\left(\mathbf{r}\right)$ and
$\Omega_{2}\left(\mathbf{r}\right)$. The dark state is not featured in $\hat{V}$, so it is not affected
by the atom-light coupling and is characterized by zero eigenenergy,
i.e., $\hat{V}|\tilde{D}\rangle=0$.

The dark and bright states can be defined up to a (position-dependent)
phase factor. The use of a different set of dark and bright states
corresponds to another gauge choice. It is convenient to use the dark
and bright states $\left|D\right\rangle $ and $\left|B\right\rangle $
expressed in terms of a relative Rabi frequency $\zeta=\Omega_{1}/\Omega_{2}$:
\begin{equation}
\left|D\right\rangle \equiv\frac{\left|\Omega_{2}\right|}{\Omega_{2}}\left|\tilde{D}\right\rangle =\frac{\left|1\right\rangle -\zeta\left|2\right\rangle }{\sqrt{1+|\zeta|^{2}}}\,,\label{eq:D-alt}
\end{equation}
\begin{equation}
\left|B\right\rangle \equiv\frac{\left|\Omega_{2}\right|}{\Omega_{2}^{*}}\left|\tilde{B}\right\rangle =\frac{1}{\sqrt{1+|\zeta|^{2}}}\left(\zeta^{*}\left|1\right\rangle +\left|2\right\rangle \right)\,.\label{eq:B-alt}
\end{equation}
In the present paper we will concentrate of the situation where $\Omega_{2}$
is real, so the two sets of dark and bright states coincide up to
a sign of $\Omega_{2}$.

For a sufficiently large single photon detuning, $\Delta\gg\Omega$,
one can adiabatically eliminate the excited state $\left|e\right\rangle $.
Specifically, applying the Schrieffer-Wolff transformation  (see e.g.
the Supplementary material of ref.~\cite{Yanes22PRL}), the effective
Hamiltonian describing the atomic dynamics in the ground state manifold
is

\begin{equation}
\hat{V}_{{\rm eff}}=\frac{V_{0}}{4}\left|\tilde{B}\right\rangle \left\langle \tilde{B}\right|=\frac{V_{0}}{4}\left|B\right\rangle \left\langle B\right|\,,\label{eq:V-eff}
\end{equation}
 with

\begin{equation}
V_{0}=\frac{\Omega^{2}}{\Delta+i\Gamma/2}\,,\label{eq:V_0}
\end{equation}
 where the real and imaginary parts of $V_{0}$ describe, respectively,
the light shift and the decay of the bright state. The dark state
is characterized by zero eigenenergy also in the projected description
$\hat{V}_{{\rm eff}}|\tilde{D}\rangle=0$, so the dark state neither decays, nor experiences the light shift, like in the full description involving all three atomic internal states.  

\section{Adiabatic motion in the dark state manifold\label{sec:Adiabatic-motion-in}}

Including the kinetic energy, the combined internal and center of
mass atomic dynamics is described by the Hamiltonian: 
\begin{equation}
\hat{H}=\frac{\hat{\mathbf{p}}^{2}}{2M}+\hat{V}\,,\label{eq:H_full}
\end{equation}
where $\mathbf{p}=-i\hbar\mathbf{\nabla}$ is the momentum operator,
$M$ is the atomic mass and $\hat{V}$ is the atom-light coupling
Hamiltonian given by Eqs.~(\ref{eq:V-definition}) or (\ref{eq:V-definition-1}).

The atomic combined internal and center of mass dynamics is generally
described by a multi-component state-vector $\left|\psi\left(\mathbf{r}\right)\right\rangle $
which can be cast in terms of bare atomic internal states $\left|1\right\rangle $,
$\left|2\right\rangle $ and $\left|e\right\rangle $, or alternatively,
in the basis of dark, bright and excited states $\left|D\right\rangle \equiv\left|D\left(\mathbf{r}\right)\right\rangle $,
$\left|B\right\rangle \equiv\left|D\left(\mathbf{r}\right)\right\rangle $
and $\left|e\right\rangle $: \textbf{}
\begin{equation}
\left|\psi\left(\mathbf{r}\right)\right\rangle =\psi_{D}\left(\mathbf{r}\right)\left|D\left(\mathbf{r}\right)\right\rangle +\psi_{B}\left(\mathbf{r}\right)\left|B\left(\mathbf{r}\right)\right\rangle +\psi_{e}\left(\mathbf{r}\right)\left|e\right\rangle \,,\label{eq:|psi>-expansion-dark, bright, excited}
\end{equation}
where $\psi_{D}\left(\mathbf{r}\right)$, $\psi_{B}\left(\mathbf{r}\right)$
and $\psi_{e}\left(\mathbf{r}\right)$ are wavefunctions for the
center of motion of atoms in the corresponding internal states. For
a sufficiently large detuning, $\Delta\gg\Omega$, one can consider
the state vector projected onto the dark and bright state manifolds
$\left|\psi_{\mathrm{eff}}\left(\mathbf{r}\right)\right\rangle =\psi_{D}\left(\mathbf{r}\right)\left|D\left(\mathbf{r}\right)\right\rangle +\psi_{B}\left(\mathbf{r}\right)\left|B\left(\mathbf{r}\right)\right\rangle $.
The evolution of $\left|\psi_{\mathrm{eff}}\left(\mathbf{r}\right)\right\rangle $
is then governed by the effective Hamiltonian $\hat{H}_{{\rm eff}}$
which has the form of Eq.~(\ref{eq:H_full}) with the atom-light
operator $\hat{V}$ replaced by the effective operator $\hat{V}_{{\rm eff}}$
given by Eq.~(\ref{eq:V-eff}). Such an approach has been applied
in refs.~\cite{Dum96PRL,Dalibard24arXiv} (with the real valued coupling
strength $V_{0}$ corresponding to neglection of losses)\textcolor{black}{~\footnote{This applies to the arXiv version of ref.~\cite{Dalibard24arXiv}, which was available during the original submission of our manuscript to Physical Review Research and to arXiv. Subsequently, a lossy excited level of the Lambda scheme was added in the “End Matter” of the revised ref.~\cite{Dalibard24arXiv}.}}. Our numerical
analysis will be based on the dynamics of the complete state-vector
$\left|\psi\left(\mathbf{r}\right)\right\rangle $ governed by
the full Hamiltonian $\hat{H}$ of Eq.~(\ref{eq:H_full}), and a
comparison to the projected dynamics described by $\hat{H}_{{\rm eff}}$
will be discussed.  We will see that in the regime where the atoms
adiabatically follow the dark state, the complete and projected dynamics
governed by the Hamiltonian $\hat{H}$ and $\hat{H}_{{\rm eff}}$
yield very similar energy spectra (especially for lower energies)
even for moderate detuning $\Delta$. Interestingly, the effects of
losses are also similar using the two approaches,  and the reason
for this will be elucidated in Sec.~\ref{subsec:Spectrum}. 

\subsection{Adiabatic approximation}

When the total Rabi frequency $\Omega$ greatly exceeds the characteristic
energy of the atomic center of mass motion, the atoms adiabatically
follow the lossless dark state to a good approximation. \textbf{}
The atomic center\textbf{} of mass motion can then be described by the
dark-state wave-function $\psi_{D}\left(\mathbf{r}\right)$ whose
evolution is governed by the following Hamiltonian \cite{Juz05PRA,Juz05JPB,Dalibard11RMP,Goldman2014}

\begin{equation}
\hat{H}_{{\rm D}}=\frac{1}{2M}(-i\hbar\mathbf{\nabla}-\mathbf{A})^{2}+\phi\,,\label{eq:H_D}
\end{equation}
where $\mathbf{A}\equiv\mathbf{A}(\mathbf{r})$ and $\phi\equiv\phi(\mathbf{r})$
are geometric vector and scalar potentials  
\begin{equation}
\mathbf{A}=i\hbar\left\langle D\right|\mathbf{\nabla}\left|D\right\rangle \quad\mathrm{and}\quad\phi=\frac{1}{2m}\mathbf{A}_{{\rm DB}}\cdot\mathbf{A}_{{\rm BD}}\,,\label{eq:A,phi}
\end{equation}
with 
\begin{equation}
\mathbf{A}_{{\rm DB}}=i\hbar\left\langle D\right|\mathbf{\nabla}\left|B\right\rangle \quad\mathrm{and}\quad\mathbf{A}_{{\rm BD}}=\mathbf{A}_{{\rm DB}}^{*}\label{eq:A_BD}
\end{equation}
being the off-diagonal matrix elements of the vector potential associated
with the non-adiabatic transitions between the dark and bright states.
Note that the Hamiltonian [Eq. (\ref{eq:H_D})] does not contain the dark
state energy, as the latter equals to zero.  Thus the adiabatic following
of the dark state is not accompanied with any additional scalar potential
due to the position-dependence of the dark state energy, which facilitates
simulation of the situation close to the Quantum Hall physics.

The geometric vector and scalar potentials acting on atoms the dark
state $\left|D\right\rangle $, as well as the corresponding magnetic
field $\mathbf{B}=\mathbf{\nabla}\times\mathbf{A}$ read \cite{Juz05PRA,Juz05JPB}
\begin{equation}
\mathbf{A}=\frac{i\hbar}{2}\frac{\zeta^{*}\mathbf{\nabla}\zeta-\zeta\mathbf{\nabla}\zeta^{*}}{1+|\zeta|^{2}}=-\hbar\frac{\eta^{2}}{1+\eta^{2}}\mathbf{\nabla}S\,,\label{eq:A-eff-D}
\end{equation}
\begin{equation}
\phi=\frac{\hbar^{2}}{2m}\frac{\mathbf{\nabla}\zeta^{*}\cdot\mathbf{\nabla}\zeta}{\left(1+|\zeta|^{2}\right)^{2}}=\frac{\hbar^{2}}{2m}\frac{\eta^{2}(\mathbf{\nabla}S)^{2}+(\mathbf{\nabla}\eta)^{2}}{\left(1+\eta^{2}\right)^{2}}\,,\label{eq:phi-eff-D}
\end{equation}
and 
\begin{equation}
\mathbf{B}=i\hbar\frac{1}{\left(1+|\zeta|^{2}\right)^{2}}\mathbf{\nabla}\zeta^{*}\times\mathbf{\nabla}\zeta=\hbar\frac{(\mathbf{\nabla}S)\times\mathbf{\nabla}\eta^{2}}{\left(1+\eta^{2}\right)^{2}}\,,\label{eq:B-eff-D}
\end{equation}
where $\eta$ and $S$ are, respectively, the real valued amplitude
and the phase of the relative Rabi frequency   
\begin{equation}
\zeta=\Omega_{1}/\Omega_{2}=\eta e^{iS}\,,\label{eq:zeta-definition}
\end{equation}
 where $\eta$ can be positive or negative.

\subsection{Magnetic flux for periodic atom-light coupling}

In what follows the Rabi frequencies $\Omega_{1}$ and $\Omega_{2}$
and thus the atom-light coupling $\hat{V}$ are considered to be spatially
periodic in the $\mathbf{e}_{x}$-$\mathbf{e}_{y}$ plane 
\begin{equation}
\Omega_{1,2}(\mathbf{r}+\mathbf{r}_{n,m})=\Omega_{1,2}(\mathbf{r})\,\quad\mathrm{and}\quad\hat{V}(\mathbf{r}+\mathbf{r}_{n,m})=\hat{V}(\mathbf{r})\,,\label{eq:Omega-periodic}
\end{equation}
with $\mathbf{r}_{n,m}=n\mathbf{a}_{1}+m\mathbf{a}_{2}$, where $\mathbf{a}_{1}$
and $\mathbf{a}_{2}$ are primitive vectors defining the 2D lattice,
$n$ and $m$ being integers.

Let us first consider the general properties of the vector potential
$\mathbf{A}=\mathbf{A}(\mathbf{r})$ and the corresponding magnetic
field $\mathbf{B}=\mathbf{B}(\mathbf{r})$ for atoms adiabatically
following a certain dressed state representing a non-degenerate eigenstate
of a spatially periodic atom-light coupling operator $\hat{V}(\mathbf{r})$.
In the present situation such a dressed state represents the dark
state $\left|D\right\rangle =\left|D\left(\mathbf{r}\right)\right\rangle $,
but the same arguments apply to atoms adiabatically following any
non-degenerate dressed state corresponding to any spatially periodic
atom-light coupling \cite{Juz-Spielm2012NJP}. Atomic dressed states
have the lattice periodicity, so the corresponding geometric vector
potential is also periodic: $\mathbf{A}(\mathbf{r}+\mathbf{r}_{n,m})=\mathbf{A}(\mathbf{r})$.
For this reason the total magnetic flux over an elementary cell is
zero: 
\begin{align}
\alpha & =\oint_{{\rm cell}}\!\mathbf{A}\cdot d\mathbf{r}=0.\label{eq:alpha-flux}
\end{align}
Using the Stokes theorem, the total magnetic flux $\alpha$ can be
represented as: 
\begin{align}
\alpha & =\alpha^{\prime}+\sum\oint_{{\rm singul}}\!\mathbf{A}\cdot d\mathbf{r}\,,\label{eq:alpha-flux-1}
\end{align}
where the first term is the flux due to the actual (continuous) magnetic
field $\mathbf{B}=\nabla\times\mathbf{A}$: 
\begin{equation}
\alpha^{\prime}=\iint_{{\rm cell}}\!\mathbf{B}\cdot d\mathbf{S}\,.\label{eq:alpha-prime-flux}
\end{equation}
The second term in Eq.~(\ref{eq:alpha-flux-1}) involves integration
around all (gauge-dependent) singular points of the vector potential.
From Eqs.~(\ref{eq:alpha-flux})-(\ref{eq:alpha-flux-1}) it follows
that the actual magnetic flux $\alpha^{\prime}$ over the elementary
cell is determined by the singularities of the vector potential \cite{Juz-Spielm2012NJP,Dalibard24arXiv}:
\begin{equation}
\alpha^{\prime}=-\sum\oint_{{\rm singul}}\!\mathbf{A}\cdot d\mathbf{r}\,.\label{eq:alpha-prime-flux-general-result}
\end{equation}
Note that, if the atom-light coupling involves two atomic internal
states, the eigenstates of such a two-level system (the dressed states)
can be parameterized by the azimuthal and polar angles of the Bloch
sphere. Such dressed states are not single-valued in the south or
north poles of the Bloch sphere, at which the vector potential for
the adiabatic motion in one of the dressed states can be singular
\cite{Dalibard11RMP}. By properly choosing the atom-light coupling,
one can thus get a non-zero magnetic flux over a plaquette $\alpha^{\prime}$
equal to an integer number of the Dirac quanta $2\pi\hbar$ \cite{Cooper11PRL,Juz-Spielm2012NJP,Cooper-Dalibard13PRL}. The non-zero magnetic flux over a plaquette $\alpha^{\prime}$ can also
be created for a more general situation of any spin or quasi-spin
(not necessarily quasi-spin $1/2$) adiabatically following an effective
magnetic field \cite{Juz-Spielm2012NJP}.
%\footnote{The non-zero magnetic flux over a plaquette $\alpha^{\prime}$ can be created also for a more general situation of any spin or quasi-spin (not necessarily quasi-spin $1/2$) adiabatically following an effective magnetic field \cite{Juz-Spielm2012NJP}.}

The situation is different for atoms adiabatically following the dark
state of the $\Lambda$ scheme. In fact, the dark state $\left|\tilde{D}\right\rangle $
given by Eq.~(\ref{eq:D}) is fully determined by the Rabi frequencies
$\Omega_{1}$ and $\Omega_{2}$ and thus is single valued, as long
as the total Rabi frequency $\Omega$ is non-zero, i.e. as long as
the dark state is not degenerate with the bright state and thus the
adiabatic concept is applicable. As a result, atoms adiabatically
following the dark state are affected by the vector potential which
does not have singularities and thus the magnetic flux over a plaquette
is zero: 
\begin{equation}
\alpha^{\prime}=0\,.\label{eq:alpha'}
\end{equation}
The same holds for the dark state $\left|D\right\rangle $ given by
Eq.~(\ref{eq:D-alt}), corresponding to another gauge, as the gauge
transformation can not change the measurable magnetic flux $\alpha^{\prime}.$
Thus one can not create a non-staggered magnetic flux for the dark
state atoms of the $\Lambda$ scheme within the adiabatic approach.
Yet, as we will see later, the non-staggered magnetic field can be
composed of a smooth background magnetic field and a set of narrow
tubes of a strong magnetic field with an opposite sign compensating
the background flux. \textbf{} If the flux tubes are narrow enough,
they will not influence significantly the atomic motion in the uniform
background magnetic flux. Thus one arrives effectively at a situation
close to motion of a particle in the uniform magnetic field.

\section{Atom-light coupling\label{sec:Atom-light-coupling}}

\subsection{Coupling scheme}

The situation described above can be implemented using the $\Lambda$
scheme with the following Rabi frequencies of the atom-light coupling depicted in Fig.~\ref{fig:Lambda}(b):

\begin{equation}
\Omega_{1}\left(\mathbf{r}\right)=\Omega_{0}\left[\sin\left(k\left(x+y\right)\right)-i\sin\left(k\left(x-y\right)\right)\right]\,\label{eq:Omega_1-specific}
\end{equation}
and 
\begin{equation}
\Omega_{2}\left(\mathbf{r}\right)=\epsilon\Omega_{0}\left[1+\frac{\nu}{2}\cos\left(2kx\right)+\frac{\nu}{2}\cos\left(2ky\right)\right]\,.\label{eq:Omega_2-specific}
\end{equation}
The experimental feasibility of this setup will be discussed later
in the paper.

The Rabi frequencies given by Eqs.~(\ref{eq:Omega_1-specific})-(\ref{eq:Omega_2-specific})
accommodate the schemes considered in refs.~\cite{Gvozdiovas23PRA}
and \cite{Dalibard24arXiv}. For $\nu=0$, the scheme reduces to a
situation analyzed in ref.~\cite{Gvozdiovas23PRA}. In particular
it was demonstrated, that for $\nu=0$ and $\epsilon\ll1$ one can
create a 2D Kronig-Penney lattice representing a set of 2D subwavelength
potential peaks affected by a staggered magnetic flux \cite{Gvozdiovas23PRA}.
\textbf{} On the other hand, ref.~\cite{Dalibard24arXiv} concentrated
on the situation where $\epsilon=1$ and $\nu=1$ for creating topological
bands. 

\subsection{Symmetries of Hamiltonian\label{subsec:Symmetries-of-Hamiltonian}}

The atom-light coupling operator $\hat{V}(\mathbf{r})$ characterized
by the Rabi frequencies $\Omega_{1,2}\left(\mathbf{r}\right)$,
Eqs.~(\ref{eq:Omega_1-specific})-(\ref{eq:Omega_2-specific}), is
invariant with respect to spatial shifts by the lattice vectors 
\begin{equation}
\mathbf{a}_{1}\equiv\mathbf{a}_{x}=a\mathbf{e}_{x}\quad\mathrm{and}\quad\mathbf{a}_{2}\equiv\mathbf{a}_{y}=a\mathbf{e}_{y}\,,\label{eq:a_x,y}
\end{equation}
where 
\begin{equation}
a=2\pi/k\label{eq:a}
\end{equation}
is the lattice constant, with $\mathbf{e}_{x}$ and $\mathbf{e}_{y}$
being the unit Cartesian vectors. Additionally, as in the case of
$\nu=0$ \cite{Gvozdiovas23PRA}, one has 
\begin{equation}
\begin{aligned} & \Omega_{1}(\mathbf{r}+\mathbf{a}_{x}/2)=\Omega_{1}(\mathbf{r}+\mathbf{a}_{y}/2)=-\Omega_{1}\left(\mathbf{r}\right)\,,\\
 & \Omega_{2}(\mathbf{r}+\mathbf{a}_{x}/2)=\Omega_{2}(\mathbf{r}+\mathbf{a}_{y}/2)=\Omega_{2}(\mathbf{r})\,.
\end{aligned}
\end{equation}
Hence the Hamiltonian $\hat{H}$ is invariant with regards to an operation
\begin{equation}
\hat{T}_{\mathbf{a}_{u}/2}=\hat{U}\exp\left(\frac{i\,\mathbf{a}_{u}\cdot\hat{\mathbf{p}}}{2\hbar}\right)\quad\left(u=x,y\right)\label{eq:T_a_pm}
\end{equation}
involving spatial translations by a half of the lattice vectors accompanied
by a self-inverse (involutory) unitary operation $\hat{U}$ \cite{Gvozdiovas23PRA}:
\begin{equation}
\hat{U}=\left|1\right\rangle \left\langle 1\right|-\left|2\right\rangle \left\langle 2\right|-\left|0\right\rangle \left\langle 0\right|,\quad{\rm with}\quad\hat{U}^{2}=\hat{I}\,.\label{eq:U}
\end{equation}
The dark state is not affected by such a combined shift: $\hat{T}_{x,y}\left|\tilde{D}\right\rangle =\left|\tilde{D}\right\rangle $,
so the geometric vector and scalar potentials $\mathbf{A}\left(\mathbf{r}\right)$
and $\phi\left(\mathbf{r}\right)$ have a periodicity $a/2$ equal
to half of the original lattice constant along the $x$ and $y$ axes:
\begin{equation}
\mathbf{A}\left(\mathbf{r}+\mathbf{a}_{u}/2\right)=\mathbf{A}\left(\mathbf{r}\right)\,,\quad\phi\left(\mathbf{r}+\mathbf{a}_{u}/2\right)=\phi\left(\mathbf{r}\right)
% \,,\quad\mathrm{with}\quad u=x,y.
\label{eq:A,phi-symmetry}
\end{equation}
with $u=x,y$.
As the Hamiltonian $\hat{H}$ commutes with the combined shift operators
$\hat{T}_{\mathbf{a}_{u}/2}$, these operators share a common set
of eigenstates 
\begin{equation}
\left|\psi_{s}^{(\mathbf{q})}(\mathbf{r})\right\rangle =e^{i\mathbf{q}\cdot\mathbf{r}}\left|u_{s}^{(\mathbf{q})}(\mathbf{r})\right\rangle \,,\label{eq:|psi-q>}
\end{equation}
with 
\begin{equation}
\hat{H}\left|\psi_{s}^{(\mathbf{q})}(\mathbf{r})\right\rangle =E_{s}(\mathbf{q})\left|\psi_{s}^{(\mathbf{q})}(\mathbf{r})\right\rangle \,,\label{eq_eigen_H_Bloch}
\end{equation}
where $\left|u_{s}^{(\mathbf{q})}(\mathbf{r})\right\rangle $
is a part invariant with respect to the combined shift by a half of
the lattice constant 
\begin{equation}
\hat{T}_{\mathbf{a}_{u}/2}\left|u_{s}^{(\mathbf{q})}(\mathbf{r})\right\rangle =\left|u_{s}^{(\mathbf{q})}(\mathbf{r})\right\rangle \,.\label{eq:|psi-q>-1}
\end{equation}
Here $s$ labels the Bloch bands characterized by the energy $E_{s}(\mathbf{q})$
and quasi-momentum $\mathbf{q}=q_{x}\mathbf{e}_{x}+q_{y}\mathbf{e}_{y}$,
the latter covering an extended Brillouin zone which is twice larger
in each Cartesian direction: $-2\pi/a\le q_{x,y}<2\pi/a$. In calculating
the energy spectrum of the system we will look for the eigenstates
of the form of Eq.~(\ref{eq:|psi-q>}). Note that describing the
atom-light coupling in terms of the projected operator $\hat{V}_{{\rm eff}}$
given by Eq.~(\ref{eq:V-eff}), the operator $\hat{U}$ featured
in the combined shift operator (\ref{eq:T_a_pm}) reduces to the $\sigma_{z}$,
as in ref.~\cite{Dalibard24arXiv}.

Consider next the time reversal operation $\mathcal{T}$ involving
the complex conjugation $K$ and a spatial shift by a half of the
lattice constant along both axes simultaneously: $\mathbf{r}\rightarrow\mathbf{r}+\left(\mathbf{a}_{x}+\mathbf{a}_{y}\right)/2$:

\begin{equation}
\mathcal{\mathcal{T}}=\exp\left(\frac{i\,\left(\mathbf{a}_{x}+\mathbf{a}_{y}\right)\cdot\hat{\mathbf{p}}}{2\hbar}\right)K\,.\label{eq:T-time reversal}
\end{equation}
Without including the excited state decay ($\Gamma=0$), the time
reversal transformation flips the sign of the parameter $\nu$ in
the Hamiltonian $\hat{H}\equiv\hat{H}_{\nu}$: 
\begin{equation}
\mathcal{\mathcal{T}}:\quad\hat{H}_{\nu}\rightarrow\hat{H}_{-\nu}\,,\label{eq:Time reversal H_c}
\end{equation}
so the Hamiltonian $\hat{H}=\hat{H}_{\nu}$ breaks the time reversal
symmetry unless $\nu=0$. Generally 2D systems without the time reversal
symmetry are characterized by integer Chern numbers describing the
Quantum Hall response \cite{Haldane88PRL,Chiu16RMP}. Thus for $\nu\ne0$
the energy bands of the system can be characterized by non-zero Chern
numbers $c$ which reverse the sign $c\rightarrow-c$ by changing
$\nu\rightarrow-\nu$, leading to reversal of the currents in the
chiral edge states.

\subsection{Effective magnetic field and scalar potential}

\subsubsection{Plots}

\begin{figure}[tbh!]
\centering\includegraphics[width=1.0\columnwidth]{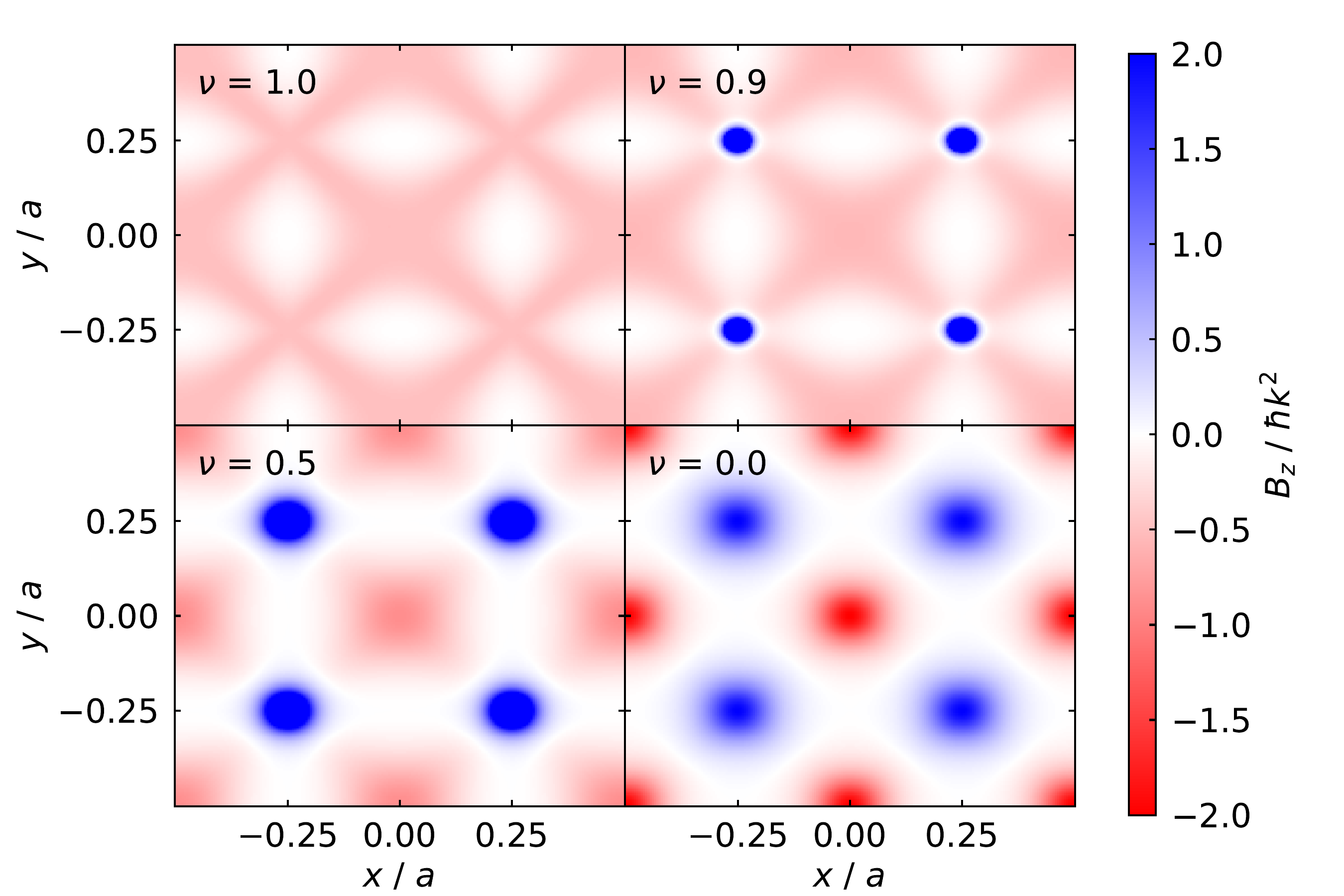}
% {\string"Article Figures/combined_phi_and_B_vs_xy!2025-05-19!01.56.09\string".png}
\caption{Spatial dependence of the 
%\textbf{(a)}
magnetic field $B_{z}$ 
% and
%\textbf{(b)} geometric potential $\phi$ 
for $\epsilon=1$ and $\nu=\left\{ 1,0.9,0.5,0\right\} $
when $\Delta=2000E_{{\rm R}}$, $\Gamma=1000E_{{\rm R}}$ and $\Omega_{0}=2000E_{{\rm R}}$.
Note that the magnetic field
%and scalar potential
around the narrow
peaks are far beyond the range of their values shown. Specifically,
the largest $\left|B_{z}\right|$ are equal approximately
to $\left\{ 1,400,16,4\right\} \hbar k^{2}$ for $\nu=\left\{ 1,0.9,0.5,0\right\} $, respectively. The geometric scalar potential $\phi$  behaves very similar to the absolute value of $B_{z}$.}   
%\db{One could add comment that even though they look similar, that away from peaks, they differ substantially. I added such a sentence further away, but it could also be mentioned here.} \db{Scalar potential visually very similar, consider removing.}

\label{fig:magnetic-field-c-grid} 
\end{figure}

\begin{figure}[tbh!]
\centering\includegraphics[width=0.75\columnwidth]{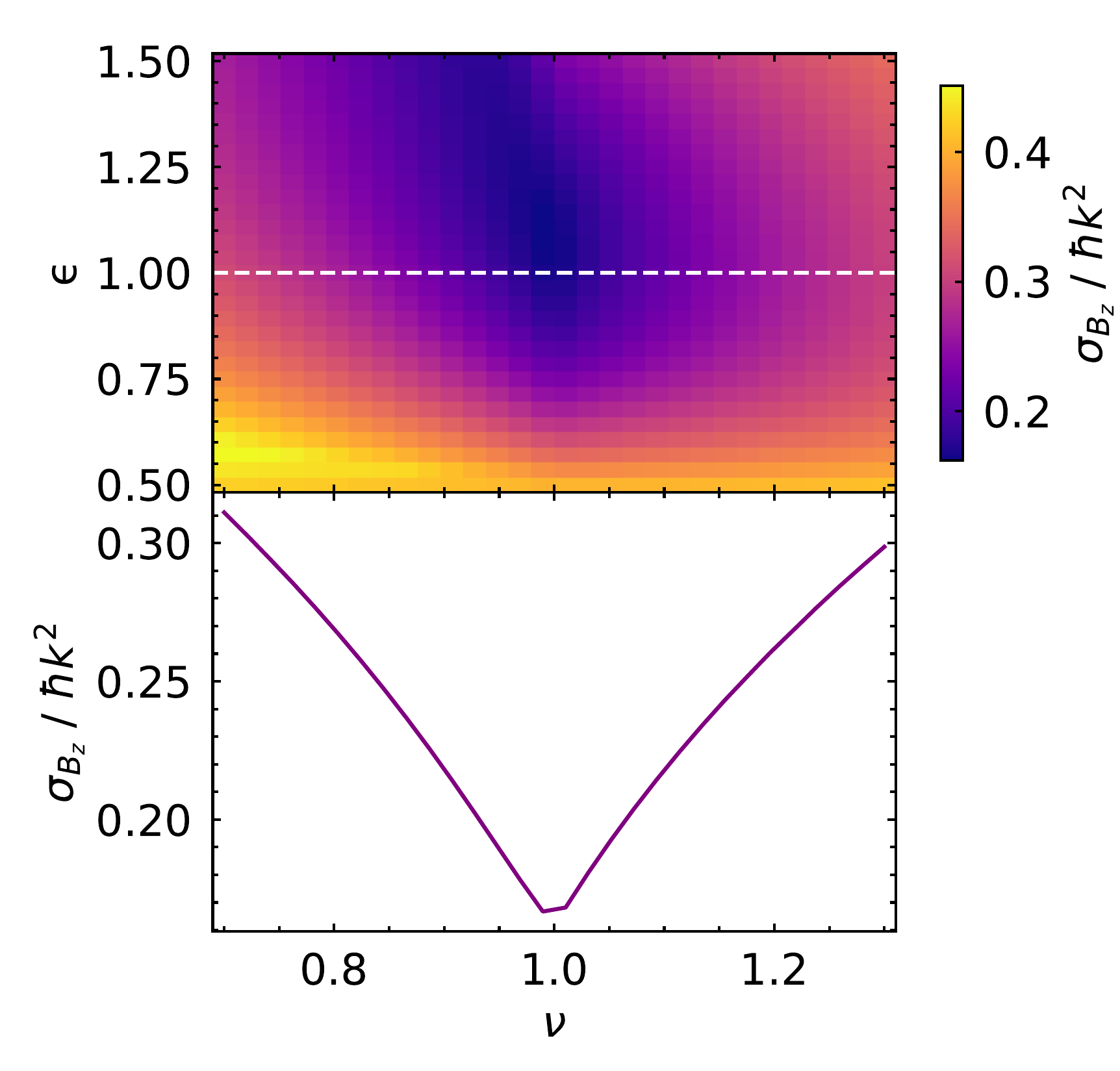}
% {\string"Article Figures/std_dev_B_z_vs_c_epsilon_with_hline!2025-05-19!01.53.21\string".png}
\caption{Dependence of the standard deviation of magnetic field $B_{z}$ on
$\epsilon$ and $\nu$ (upper plot), as well as on $\nu$ for $\epsilon=1$
(lower plot). Other parameters are $\Delta=2000E_{{\rm R}}$, $\Gamma=1000E_{{\rm R}}$,
$\Omega_{0}=2000E_{{\rm R}}$. For $\nu\protect\ne1$ we have eliminated
narrow peaks of the magnetic field by cutting the magnetic field exceeding the maximum of the absolute value of $B_{z}$ corresponding to $\epsilon=1$ and $\nu=0$. The standard deviation of the scalar potential $\phi$ is characterized
by nearly identical plots.
% \db{Should consider 1x2 instead of 2x1.}
}

\label{fig:std-dev-B-phase-diag-1} 
\end{figure}

Figure~\ref{fig:magnetic-field-c-grid} shows the amplitude $B_{z}$
of the effective magnetic field $\mathbf{B}=B_{z}\mathbf{e}_{z}$
%and (b) the geometric scalar potential $\phi$
for $\epsilon=1$ and
different values of $\nu\ge0$. For $\nu=0$ the positive and negative
values of $B_{z}$ are distributed in the same way, but are shifted
with respect to each other by $(\mathbf{a}_{x}+\mathbf{a}_{y})/2$
or $(\mathbf{a}_{x}-\mathbf{a}_{y})/2$, i.e. by half of a lattice
constant $a$ simultaneously along both Cartesian axes $x$ and $y$.
Increasing $\nu$, the negative magnetic field spreads out, whereas
the positive magnetic field becomes concentrated in smaller spots
with a larger strength, so that the total magnetic flux over an elementary
cell sums to zero, as required by Eq.~(\ref{eq:alpha'}). Note that
for $\nu\rightarrow\pm1$, the flux tubes become infinitely
narrow and thus reduce to the 2D Dirac delta functions which are not
visible in Fig.~\ref{fig:magnetic-field-c-grid} for $\epsilon=\nu=1$. Outside the areas of the narrow peaks, the magnetic field is the
most uniform for $\epsilon=1$ and $\nu\rightarrow1$, as one can
see in Fig.~\ref{fig:std-dev-B-phase-diag-1}. \textbf{ } Reversing
the sign of $\nu$, the magnetic field remains distributed the same
as in Fig.~\ref{fig:magnetic-field-c-grid} subject to the spatial
shift by $(\mathbf{a}_{x}+\mathbf{a}_{y})/2$ or $(\mathbf{a}_{x}-\mathbf{a}_{y})/2$,
as well as the reversal of the sign of the magnetic field. Therefore
in what follows we will concentrate mostly on the situation where
$\nu\ge0$. The geometric scalar potential $\phi$ 
%plotted in Fig.~\ref{fig:magnetic-field-c-grid}(b)
behaves very similarly to the absolute value of the magnetic field
$\left|B_{z}\right|$ and has the same kind of narrow peaks, as
follows from Eqs.~(\ref{eq:B_z,phi_small-rho}) and (\ref{eq:f}) presented
below. 
%Furthermore, the standard deviation of the geometric scalar potential $\phi$ is very similar to the one shown in Fig.~\ref{fig:std-dev-B-phase-diag-1} for $B_{z}$, and the scalar potential is also the most uniform for $\epsilon=1$ and $\nu\rightarrow1$. 

\subsubsection{Analysis in a vicinity of zero points of $\Omega_{1}\left(\mathbf{r}\right)$}

\begin{figure}[tbh!]
\centering\includegraphics[width=0.95\columnwidth]{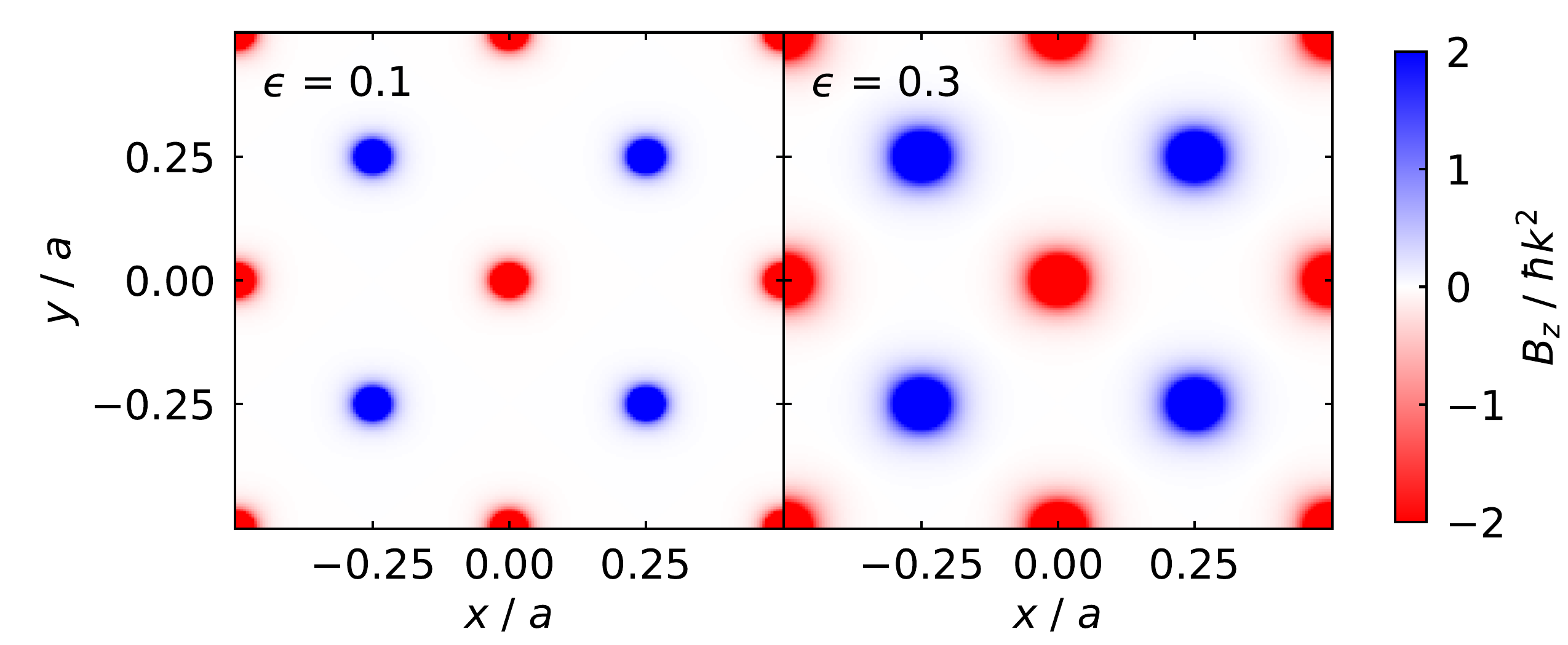}
% {\string"Article Figures/combined_phi_and_B_vs_xy_v2!2025-05-19!01.52.14\string".png}
\caption{Spatial dependence of the magnetic field $B_{z}$ 
%and \textbf{(b)} geometric potential $\phi$
for $\nu=0$ and $\epsilon=\left\{ 0.1,0.3\right\} $
when $\Delta=2000E_{{\rm R}}$, $\Gamma=1000E_{{\rm R}}$ and $\Omega_{0}=2000E_{{\rm R}}$.
Note that the magnetic field around the narrow
peaks are far beyond the range of their values shown. Specifically,
the largest $\left|B_{z}\right|$ are equal approximately
to $\left\{ 400,44.4444\right\} \hbar k^{2}$
%and $\left\{ 400,44.4444\right\} E_{{\rm R}}$
for $\epsilon=\left\{ 0.1,0.3\right\} $.
The geometric scalar potential $\phi$  looks very similar to the absolute value of $B_{z}$.}

\label{fig:magnetic-field-epsilon-grid-1} 
\end{figure}

The appearance of narrow magnetic flux tubes featured in Fig.~\ref{fig:magnetic-field-c-grid}
as $\nu$ is close to the unity, can be understood from the following
arguments. The Rabi frequency $\Omega_{1}\left(\mathbf{r}\right)$
given by Eq.~(\ref{eq:Omega_1-specific}) equals to zero when $x+y=na/2$
and $x-y=ma/2$, where $n$ and $m$ are integers. Around these zero
points of $\Omega_{1}\left(\mathbf{r}\right)$, one can write
$x+y=na/2+\rho\cos\varphi$ and $x-y=ma/2+\rho\sin\varphi$ with $\rho\ll a$.
Thus the relative Rabi frequency $\zeta$ given by Eq.~(\ref{eq:zeta-definition})
reads up to the leading terms in the radial deviation $\rho$:

\begin{equation}
\zeta\approx\eta_{\rho}e^{\pm i\varphi}\,,\quad\mathrm{with}\quad\eta_{\rho}=\left(-1\right)^{n}\frac{k\rho}{\epsilon\left(1\pm\nu\mp2ck^{2}\rho^{2}\right)}\,,\label{eq:zeta-n,m}
\end{equation}
where the sign  $\pm 1 =\left(-1\right)^{n+m}$ 
%corresponds to the sign of 
%$\left(-1\right)^{n+m}$  
 corresponds to the phase winding of the Rabi frequency $\Omega_{1}\left(\mathbf{r}\right)$,  represented by arrows in Fig.~\ref{fig:Lambda}(b).
In the denominator of $\eta_{\rho}$, we have retained a term quadratic
in $\rho$, because the constant term $1\pm\nu$ becomes vanishingly
small as $\nu\rightarrow\mp1$. Using Eq.~(\ref{eq:A-eff-D}) for
$\mathbf{A}$ with $\zeta$ given by Eq.~(\ref{eq:zeta-n,m}), the
magnetic flux over the radius $\rho$ is 
\begin{equation}
\alpha_{\rho}=-\oint_{\rho}\!\mathbf{A}\cdot d\mathbf{r}\,=\mp2\pi\hbar\frac{\eta_{\rho}^{2}}{1+\eta_{\rho}^{2}}\,.\label{eq:alpha_rho}
\end{equation}
If $\nu\ne\mp1$, the magnetic flux $\alpha_{\rho}$ is zero for $\rho=0$.
Subsequently, the flux increases with $\rho$ and approaches the maximum
flux value equal to $\mp2\pi\hbar$ as $\eta_{\rho}^{2}\gg1$, where
$2\pi\hbar$ is the Dirac flux quantum. A characteristic radius $\rho_{0}$
over which the magnetic field is concentrated can be estimated by
taking $\eta_{\rho_{0}}^{2}\approx1$, which gives for $\epsilon\left|1\pm\nu\right|\ll1$:
\begin{equation}
k\rho_{0}=\epsilon\left|1\pm\nu\right|\,.\label{eq:k rho_0}
\end{equation}
Consider next the synthetic magnetic field and the geometric scalar potential
around the zero points of $\Omega_{1}\left(\mathbf{r}\right)$ with $\epsilon\left|1\pm\nu\right|\ll1$, where   $|\Omega_{2}\left(\mathbf{r}\right)|$ is minimum (see Fig.~\ref{fig:Lambda}(b)), so the changes of the relative Rabi frequency $\zeta = \Omega_{1}/\Omega_{2}$ generating the geometric potentials are maximum.
Substituting Eq.~(\ref{eq:zeta-n,m}) into Eqs.~(\ref{eq:phi-eff-D}) and (\ref{eq:B-eff-D}), one has:
% and (\ref{eq:B-eff-D}), one has \db{Seems to be factor of 2 too large in numerics, why?} \db{Not sure about +- sign for $B_z$.}
\begin{equation}
B_{z}=\mp2\hbar k^{2}f\left(\rho\right)\,,\quad\phi=2E_{\mathrm{R}}f\left(\rho\right)\,,\label{eq:B_z,phi_small-rho}
\end{equation}
where 
\begin{equation}
E_{\mathrm{R}}=\frac{\hbar^{2}k^{2}}{2m}\label{eq:E_rec}
\end{equation}
is the recoil energy and 
\begin{equation}
f\left(\rho\right)=\frac{k^{2}\rho_{0}^{2}}{\left(k^{2}\rho^{2}+k^{2}\rho_{0}^{2}\right)^{2}}\,,\label{eq:f}
\end{equation}
with $\rho_{0}$ as in Eq.~(\ref{eq:k rho_0}). Thus close to the zero points of $\Omega_1(\mathbf{r})$  with $\epsilon\left|1\pm\nu\right|\ll1$, the effective magnetic field $B_{z}$ and geometric scalar potential
$\phi$ have the spatial dependence characterized by the same function
$f\left(\rho\right)$, like the geometric potentials produced
by vortex light beams \cite{Braver2025}.
As $k\rho_{0}\ll1$, the
function $f\left(\rho\right)$ describes narrow and large peaks of
the magnetic field and the scalar potential with the radius $\rho_{0}=\epsilon\left|1\pm\nu\right|/k$
much smaller than the lattice constant $a=2\pi/k$. For example, if
$\nu>1$, the condition $\epsilon\left|1\pm\nu\right|\ll1$
is fulfilled for the lower sign (odd $n+m$) corresponding to the minima of $|\Omega_{2}\left(\mathbf{r}\right)|$ in Fig.~\ref{fig:Lambda}(b). This provides narrow positive peaks
of the magnetic field at the zero points of $\Omega_{1}\left(\mathbf{r}\right)$
with odd $n+m$, as one can see in Fig.~\ref{fig:magnetic-field-c-grid}.
On the other hand, if $\epsilon\ll1$, the narrow peaks of the magnetic
field and the scalar potential can be formed also for the time-reversal
case with $\nu=0$ \cite{Gvozdiovas23PRA}. In such a situation, the
narrow magnetic fluxes proportional to $\mp1=\left(-1\right)^{n+m+1}$
alternate their sign depending on the parity of $n+m$, and there
is no smooth background magnetic flux, as illustrated in Fig.~\ref{fig:magnetic-field-epsilon-grid-1}.
It is instructive that width of the function $f\left(\rho\right)$
defining the magnetic field and the geometric scalar potential in
Eqs.~(\ref{eq:B_z,phi_small-rho})-(\ref{eq:f}), is proportional
to $\rho_{0}$ whereas the height of the peaks goes as $1/\rho_{0}^{2}$.
Therefore the function $f\left(\rho\right)$ becomes proportional
to the 2D Dirac delta function as $k\rho_{0}\rightarrow0$. Note that
for 1D dark-state lattices one can also produce the subwavelength
potential barriers of the width $x_{0}$ and the height proportional
$1/x_{0}^{2}$ \cite{Zoller2016,Jendrzejewski2016}. Yet the area
over such a 1D barrier is proportional to $1/x_{0}$ and thus increases
to infinity as $x_{0}$ goes to zero, making the tunneling over the
1D subwavelength dark-state barriers fully suppressed.  

For $\nu=\mp1$, Eq.~(\ref{eq:zeta-n,m}) yields $\eta_{\rho}\propto1/\rho$,
so the magnetic flux $\alpha_{\rho}$ given by Eq.~(\ref{eq:alpha_rho})
reaches a finite value $\pm2\pi\hbar$ as $\rho\rightarrow0$. Hence,
the narrow magnetic fluxes, featured for $\epsilon\left|1\pm\nu\right|\ll1$,
reduce to the Aharonov-Bohm type singularities as $\nu=\mp1$. Yet,
in practice one can not have the perfect tuning; even a tiny deviation
from $\nu=\mp1$ makes the magnetic flux somewhat extended around
$\rho=0$, removing the perfect Aharonov-Bohm singularity. On the
other hand, if $\nu=\mp1$ and hence $\rho_{0}=0$, the dark and bright
states are degenerate at $\rho=0$. Strictly speaking, in this region
the adiabatic approach is no longer applicable and one can not use
the concept of the geometric vector potential and magnetic field,
as well as the geometric scalar potential. Let us consider this issue
in more detail. 

\subsubsection{Applicability of adiabaticity}

The applicability of adiabaticity can be characterized by the ratio
$\gamma\left(\rho\right)$ between the scalar potential $\phi$ (describing
the non-adiabatic transitions) and the energy difference between the
dark and bright states $E_{\mathrm{BD}}\approx\Omega^{2}/\Delta$, with $\Omega \equiv \Omega\left(\mathbf{r}\right)$.
In a vicinity of $\rho=0$, the energy difference reads 
\begin{equation}
E_{\mathrm{BD}}\approx\Omega_{0}^{2}k^{2}\left(\rho^{2}+\rho_{0}^{2}\right)/\Delta\,.\label{eq:E_DB-1}
\end{equation}
Using also Eqs.~(\ref{eq:B_z,phi_small-rho})-(\ref{eq:f}) for $\phi$,
the ratio $\gamma\left(\rho\right)$ is 
\begin{equation}
\gamma\left(\rho\right)=\frac{\phi}{E_{\mathrm{BD}}}\approx\frac{\rho_{c1}^{4}\rho_{0}^{2}}{\left(\rho^{2}+\rho_{0}^{2}\right)^{3}}\,,\label{eq:gamma}
\end{equation}
with 
\begin{equation}
k^{4}\rho_{c1}^{4}=\frac{2E_{\mathrm{R}}\Delta}{\Omega_{0}^{2}}\,,\label{eq:rho_c1}
\end{equation}
where $k\rho_{c1}\ll1$, as the recoil energy $E_{\mathrm{R}}$ is
usually several orders of magnitude smaller than the Rabi frequency
$\Omega_{0}$, while the detuning $\Delta$ can be of the order of
$\Omega_{0}$. The ratio $\gamma\left(\rho\right)$ is maximum for
$\rho=0$, where
\begin{equation}
\left.\gamma\left(\rho\right)\right|_{\rho=0}=\gamma_{0}=\rho_{c1}^{4}/\rho_{0}^{4}\,.\label{eq:gamma-1}
\end{equation}
If the radius $\rho_{0}$ describing the width of the effective magnetic field and the geometric scalar potential 
is equal to $\rho_{c1}$, i.e. $\rho_{0}=\rho_{c1}$, the ratio $\gamma_{0}$
equals to the unity, and the non-adiabatic transitions become significant
in a vicinity of $\rho=0$. The distance $\rho_{0}=\rho_{c1}$ thus
represents the first border line when the non-adiabatic transitions
and losses start to be important. By further decreasing $\rho_{0}$
one reaches the point $\rho_{0}=\rho_{c2}<\rho_{c1}$ for which $\gamma\left(\rho\right)=1$
at $\rho=\rho_{0}=\rho_{c2}$. Using Eq.~(\ref{eq:gamma}), the latter
radius is 
\begin{equation}
\rho_{c2}=2^{-3/4}\rho_{c1}\approx0.6\rho_{c1}\,.\label{eq:rho_c2}
\end{equation}
In this way, reducing the $\rho_{0}$ to $\rho_{0}=\rho_{c2}$, the
non-adiabatic transitions become important for the area $\rho<\rho_{c2}$,
in which the scalar potential and the magnetic field are concentrated.
In this region the atoms no longer adiabatically follow the dark
state and thus no longer feel the scalar potential and the magnetic
field. The laser fields are then too weak to induce a sufficient splitting
between the dark and bright states and also to provide substantial losses.
In this ``gray area'' with $\rho<\rho_{c2}$, the center of mass
evolution does not alter the atomic internal state in any considerable
way. In fact, for $\rho>\rho_{0}=\rho_{c2}$ the atomic dark state
is close to the internal state $\left|2\right\rangle $, and the atoms
remain in this state also for $\rho<\rho_{c2}$ where the Rabi frequency
$\Omega$ is too small to change the atomic internal state. In the
limit where $\nu\rightarrow\mp1$ both critical radii $\rho_{c1}$
and $\rho_{c2}$ go to zero, so there is no issue with the non-adiabaticity
due to the infinitely narrow spots of the magnetic field and the vector
potential at $\rho=0$.

\subsection{Spectrum\label{subsec:Spectrum}}

\begin{figure}[tbh!]
\centering\includegraphics[width=0.95\columnwidth]{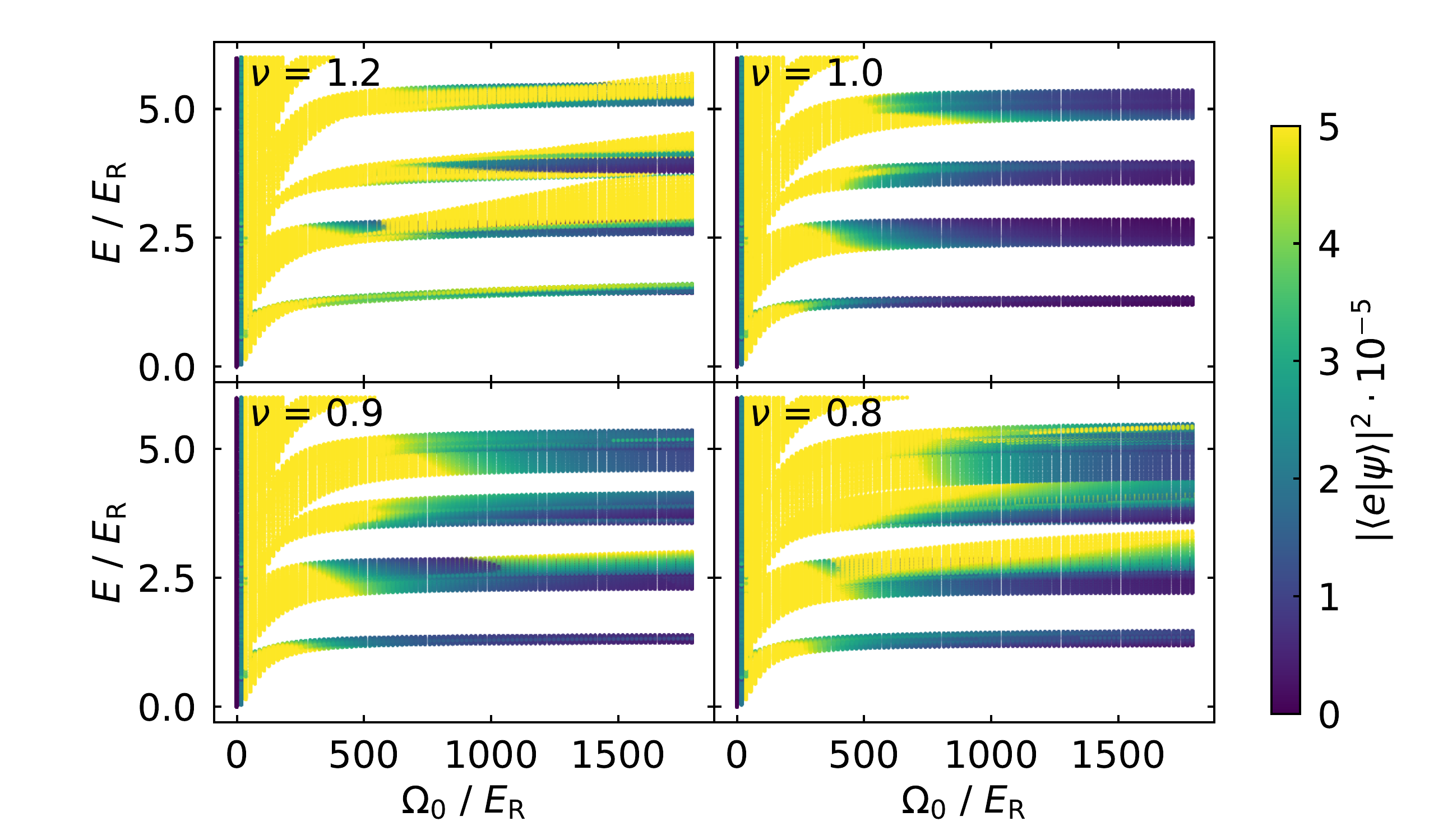}
% {\string"Article Figures/N_eigenenergies_vs_x!2025-05-19!01.50.03\string".png}
\caption{Dependence of eigenenergies $E$ on the amplitude $\Omega_{0}$ of the
Rabi frequency for the four lowest Bloch bands belonging to the dark
state manifold for $\epsilon=1$ and $\nu=\left\{ 1.2,1.0,0.9,0.8\right\} $,
with $\Delta=2000E_{{\rm R}}$ and $\Gamma=1000E_{{\rm R}}$ The calculations
have been done using the full Hamiltonian (\ref{eq:H_full}) with
the atom-light coupling operator $\hat{V}$ involving all three atomic
internal states. Colors
%of points
correspond to the excited state
population of the eigenstates $\left|\langle e|\psi\rangle\right|^{2}$ due to non-adiabatic
effects.}

\label{fig:spectrum-c-grid-1} 
\end{figure}

Figure~\ref{fig:spectrum-c-grid-1} shows the dependence of the energy
spectrum of the system on the atom-light coupling strength $\Omega_{0}$
when $\epsilon=1$ and $\nu$ is close to the unity. The plots include
the four lowest bands of the dark state manifold; the colors indicate
the population for the corresponding energy of the lossy excited state
due to non-adiabatic effects. The excited state population is proportional
to the imaginary part of the energy and is thus a measure of losses.
For $\nu=1$ the excited state population is very close to zero as
long as $\Omega_{0}$ is not too small and thus the adiabatic following
of the dark state holds to a good approximation. As $\nu$ deviates
from the unity, the excited state population increases, especially
in the upper parts of the energy spectrum. This is because departure
from $\nu=1$ leads to formation of the finite-width peaks of magnetic
fluxes and geometric scalar potential $\phi$ around the zero points
of $\Omega_{1}$ corresponding to odd $n+m$, as one can see in Fig.~\ref{fig:magnetic-field-c-grid}.
Yet, according to Eq.~(\ref{eq:A,phi}), the geometric potential
$\phi$ is proportional to $\left|\mathbf{A}_{\mathrm{DB}}\right|^{2}$,
where $\mathbf{A}_{\mathrm{DB}}$ is the off diagonal matrix element
of the vector potential responsible for the non-adiabatic coupling
between the dark and bright states. Hence the non-adiabatic coupling
is significant around the peaks of the scalar potential and magnetic
field. 

The calculations presented in Fig.~\ref{fig:spectrum-c-grid-1} have
been carried out using the complete Hamiltonian (\ref{eq:H_full}) with
the atom-light coupling operator $\hat{V}$ {[}given by Eqs.~(\ref{eq:V-definition}),
(\ref{eq:Omega_1-specific}) and (\ref{eq:Omega_2-specific}){]} involving
all three atomic internal states. We have checked that almost the
same energy spectrum and the same losses are obtained  replacing
the full atom-light coupling operator $\hat{V}$ by the projected
operator $\hat{V}_{{\rm eff}}$ {[}given by Eq.~(\ref{eq:V-eff}){]}
acting on the manifold of the dark and bright states. This can be
understand from the fact, that for a sufficiently large total Rabi
frequency $\Omega$ the atoms adiabatically follow the dark state,
so the atomic dynamics is determined exclusively by the position-dependence
of the dark state defining the geometric vector and scalar potentials
$\mathbf{A}$ and $\phi$ in Eqs.~(\ref{eq:A,phi})-(\ref{eq:A_BD}).
 In such an adiabatic dynamics, the losses originate mostly from
the spatial areas close to the zero points of the Rabi frequency $\Omega\equiv\Omega\left(\mathbf{r}\right)$
where the adiabaticity holds weaker. Yet close to its zero points
the total Rabi frequency $\Omega\left(\mathbf{r}\right)$ is so small that
the condition $\Omega\left(\mathbf{r}\right)\ll\Delta$ holds even
if the detuning $\Delta$ is equal to the Rabi frequency $\Omega_{0}$,
as it is the case for the plots presented in  Fig.~\ref{fig:spectrum-c-grid-1}. As a result, the projected description works well to describe the losses.
\begin{figure}[tbh!]
\centering\includegraphics[width=0.95\columnwidth]{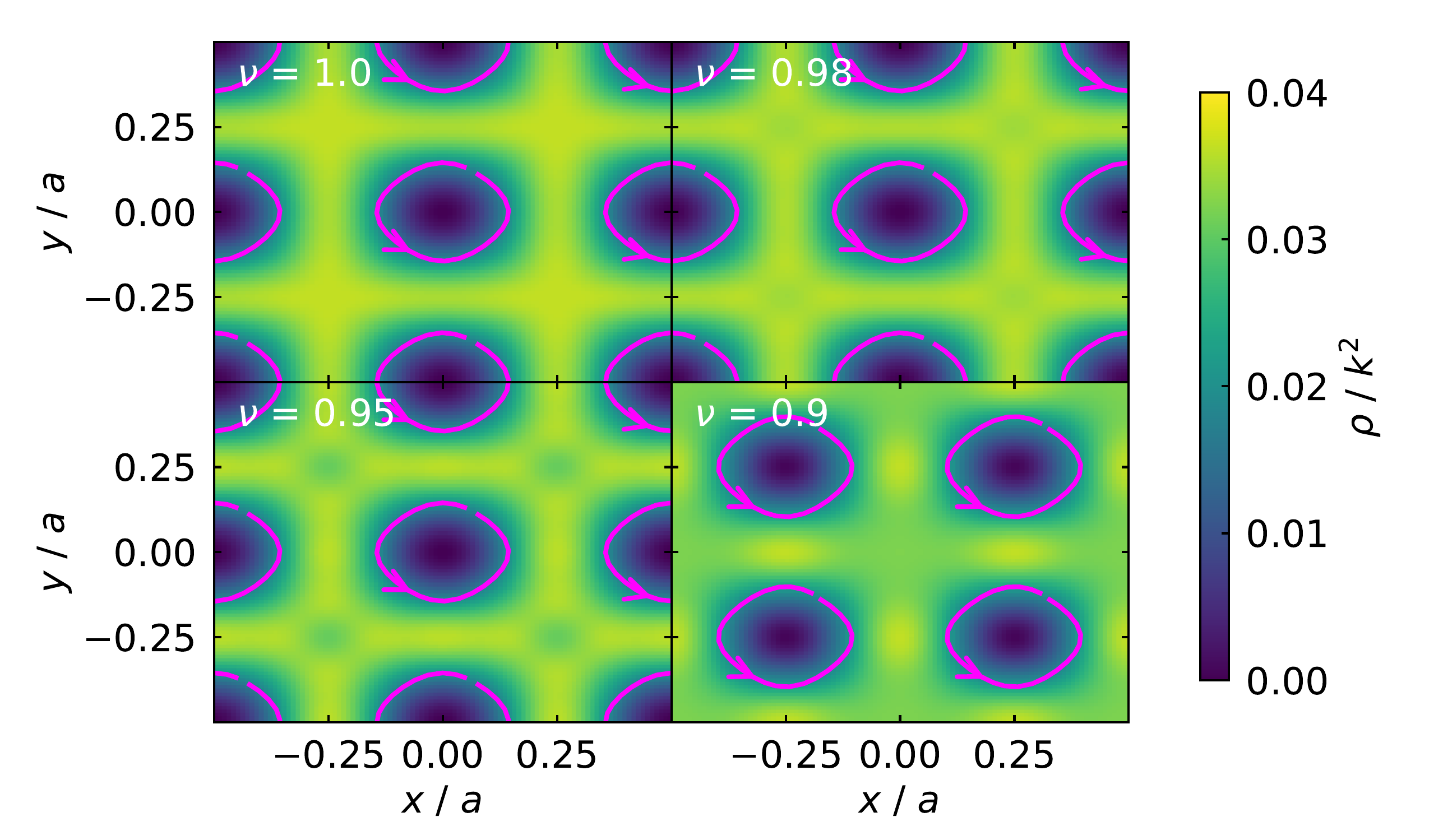}
% {\string"Article Figures/N_density_q_0_w_streamlines_vs_xy!2025-06-11!12.20.33\string".png}
\caption{Spatial dependence of the total atomic population $\rho$ in the lowest
energy Bloch state with zero quasi-momentum $\mathbf{q}=0$  for
$\epsilon=1.0$, $\nu=\left\{ 1.0,0.98,0.95,0.90\right\} $, $\Delta=2000E_{{\rm R}}$,
$\Gamma=1000E_{{\rm R}}$, $\Omega_{0}=2000E_{{\rm R}}$.}

\label{fig:total-rho-c-grid} 
\end{figure}

For $\nu$ close to the unity and large enough values of $\Omega_{0}$,
the energy bands shown in Fig.~\ref{fig:spectrum-c-grid-1} are nearly
equally spaced and have non-trivial topology characterized by unit
Chern numbers $c=1$  \footnote{We have calculated the Chern numbers $c$ of the Bloch bands using
the algorithm presented by T. Fukui, Y. Hatsugai, and H. Suzuki, Chern Numbers in Discretized Brillouin Zone: Efficient Method of Computing (Spin) Hall Conductances, J. Phys. Soc. Jpn. \textbf{74}, 1674 (2005).}, as for the perfect tuning when $\nu=1$ \cite{Dalibard24arXiv}.
The ratio of the width of individual bands to the band gap is the
smallest for $\nu=1$ and increases with the deviation from this value.
This can be understood from the fact that for $\nu\rightarrow1$ both
the magnetic field and the scalar potential are the most uniform (see
Fig.~\ref{fig:std-dev-B-phase-diag-1}), so the situation is the
closest to the Landau problem of motion of a particle in the uniform
magnetic field and uniform scalar potential, where the Bloch bands
are flat.   As illustrated in Fig.~\ref{fig:total-rho-c-grid},
for $\nu$ sufficiently close to the unity ($\nu=1\,,\,0.98$
and $0.95$), the lowest energy Bloch state with zero quasi-momentum
$\mathbf{q}=0$ has density holes centered at the zero points of $\Omega_{1}$
corresponding to even $n+m$. At these points both the smooth geometric
potential and the scalar potential are maximum (see fig.~\ref{fig:magnetic-field-c-grid}),
so it is energetically more favorable to have vortices centered there.

\begin{figure}[tbh!]
\centering\includegraphics[width=0.75\columnwidth]{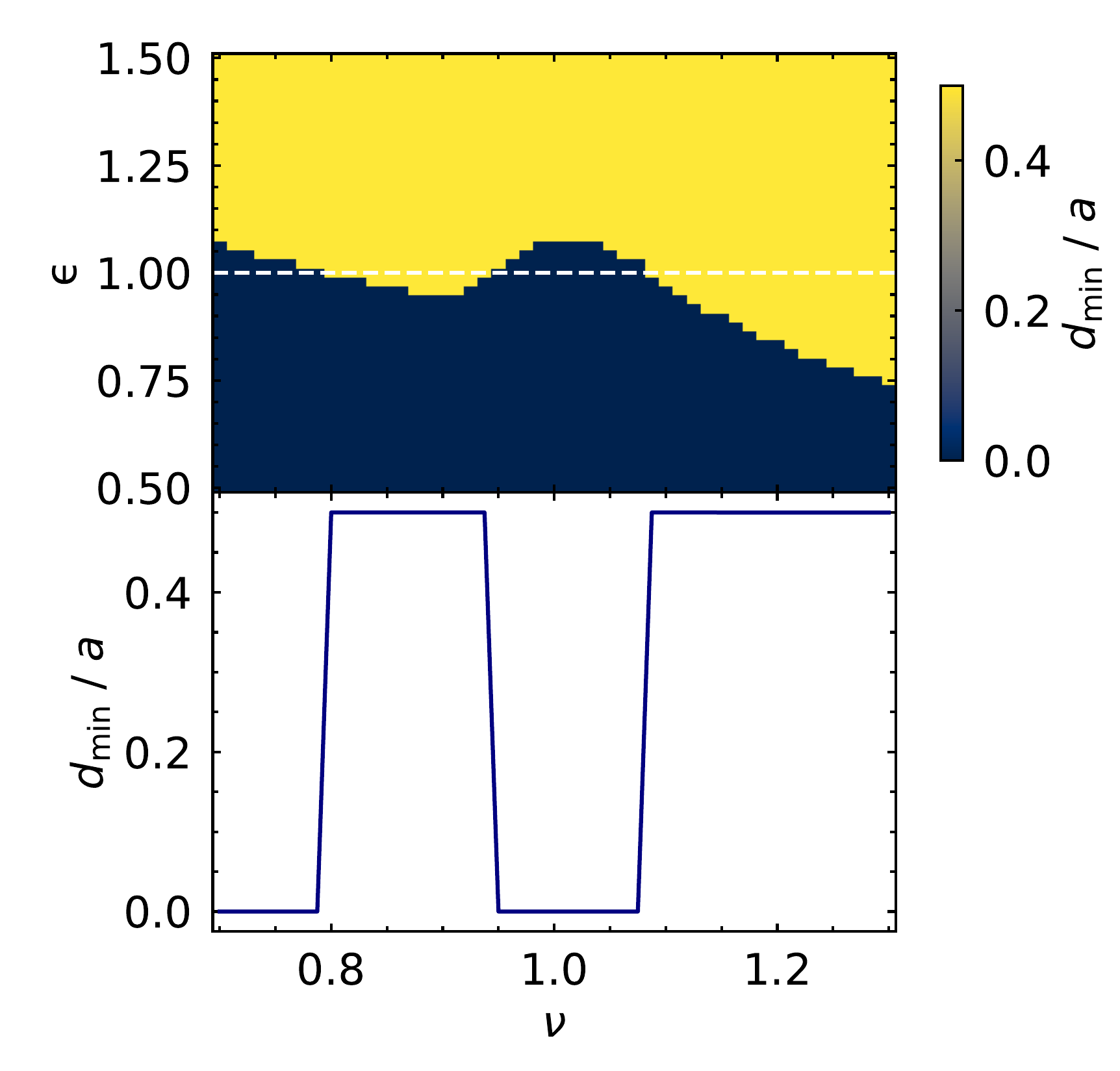}
% {\string"Article Figures/dist_rho_minimum_vs_c_epsilon_with_hline!2025-06-15!17.21.02\string".png}
\caption{
% \db{Modify colorbar with just two colors?}
Dependence on $\epsilon$ and $\nu$ of the position of the density
dips for the lowest Bloch band. Here $d_{\mathrm{min}}$ is the distance
of the density dip from $x=y=0$ corresponding to the zero point of
$\Omega_{1}$ for even $n+m$. Around $\epsilon=\nu=1$ this distance
is 0 (dark color). Beyond this area with the increase of $|1-\nu|$
the vortices become situated around the Aharonov-Bohm spikes corresponding to
the distance $d_{\mathrm{min}}$ equal to $a/2$ (light color).
}

\label{fig:d_min} 
\end{figure}

As the difference $|1-\nu|$ exceeds a critical value, it becomes more
favorable for the ground state atoms to have vortices circulating
around the narrow peaks of magnetic fluxes and the scalar potential
corresponding to the zero points of $\Omega_{1}$ with odd $n+m$.
This leads to the reorganization of the ground state taking place
for $\nu=0.9$ in Fig.~\ref{fig:total-rho-c-grid}  For the parameters
used in Fig.~\ref{fig:total-rho-c-grid}, the value $\nu=0.9$ corresponds
to the radius $\rho_{0}=\epsilon\left|1-\nu\right|/k$ equal to the
critical radius $\rho_{c2}$ given by Eqs.~(\ref{eq:rho_c1})-(\ref{eq:rho_c2}),
at which the atoms start feeling the narrow peaks of the scalar potential
and effective magnetic field.  The reorganization of the ground state
is presented in Fig.~\ref{fig:d_min}, showing how the position of the
vortex changes with $\nu$ for various values of $\epsilon$. Around
$\epsilon=\nu=1$ the vortices are located at the zero point of $\Omega_{1}$
for even $n+m$. For larger values of $1-\nu$ the vortices become
located around the Aharonov-Bohm spikes corresponding to the zero
point of $\Omega_{1}$ for odd $n+m$. By further increasing $1-\nu$,
the narrow peaks of the vector and scalar potential widen, so it becomes
unfavorable for hosting there the vortices which return to the original
position characteristic for $\nu\approx1$. Note that reorganization
of the Bloch states featured in Fig.~\ref{fig:total-rho-c-grid}
takes place before the bands touch each other and become topologically
trivial.  

The $\Lambda$ scheme with the Rabi frequencies given by Eqs.~(\ref{eq:Omega_1-specific})-(\ref{eq:Omega_2-specific})
provides the topological subwavelength optical lattice of the square
geometry. Such a scheme can be implemented if the ground states $\left|1\right\rangle $
and $\left|2\right\rangle $ are non-degenerate. Because of the energy
difference between the atomic ground states, the optical
transitions $\left|1\right\rangle \rightarrow\left|e\right\rangle $
and $\left|2\right\rangle \rightarrow\left|e\right\rangle $ can be
individually accessed by the laser fields  with the polarizations which are not necessarily well defined. The energy difference between
the ground states can be achieved using two different hyperfine ground
states or applying a strong bias magnetic field splitting the two
hyperfine states with different spin projections. The degeneracy of
the two ground states would prevent realization of this setup because
of the polarization constraint due to the transversality of the standing
electromagnetic waves and the square geometry used. Note that a similar
kind of the subwavelength topological lattice can be created using
the $\Lambda$ scheme with degenerate ground states for the triangular
geometry of the light fields \cite{Dalibard24arXiv}. In that case,
going beyond the fine tuned parameter regime considered in ref.~\cite{Dalibard24arXiv},
we have found that both the effective magnetic field and the geometric
scalar potential exhibits then narrow and intense subwavelength peaks
distributed in a hexagonal way. Away from these subwavelength spots,
magnetic field changes the sign and becomes smooth, as in the square
lattice. Topological  subwavelength lattices of non-square geometry
will be considered in more detail elsewhere.

\subsection{Existence of ideal Chern bands}

There has been a considerable recent interest in simulating a fractional
Chern insulator (FCI) using different platforms, including ultracold
atoms \cite{Cooper-Dalibard13PRL,Greiner23Nature,Joachim24PRL}. While
having a non-zero Chern number is a necessary condition for formation
of FCI, it is not sufficient to stabilize the many-body FCI phase
\cite{Wang21PRL,Crepel2023PRR,Ledwith23PRB}.  Thus the energy bands
must satisfy certain conditions, which allow a mapping to Landau levels. This can be studied via the quantum geometric tensor (QGT) $\mathcal{Q}_{\mathbf{q}}^{ab}$,
which is defined as \cite{Mitscherling2025arXiv}: 

\begin{equation}
\mathcal{Q}_{\mathbf{q}}^{ab}=\sum_{s\in\mathcal{S}}\langle\partial^{a}u_{\mathbf{q}s}|\hat{R}_{\mathbf{q}}|\partial^{b}u_{\mathbf{q}s}\rangle\,,\label{eq:Q_q^ab}
\end{equation}
% \db{Do we show Berry curvature for multiple c?}
where the partial derivatives $\partial^{a}$ and $\partial^{b}$
are taken with respect to the quasi-momentum components $q_{a}$ and
$q_{b}$, with $a,b\in\left\{ q_{x},q_{y}\right\} $. Here $|u_{\mathbf{q}s}\rangle$
is the periodic Bloch state vector at quasi-momentum $\mathbf{q}$
for the $s$-th band, $\mathcal{S}$ is the set of bands under consideration,
the operator $\hat{R}_{\mathbf{q}}=\hat{I}-\sum_{s\in\mathcal{S}}|u_{\mathbf{q}s}\rangle\langle u_{\mathbf{q}s}|$
projects onto the complement subspace and $\hat{I}$ is the identity
operator. No summation over $s$ is to be taken if we are considering
a geometric tensor just for a single Bloch band $s$. 

The form of QGT presented in Eq.~(\ref{eq:Q_q^ab}) is not convenient
for numerical calculations, as one would need a continuous phase gauge
for the Bloch state vectors. Yet, one can express the QGT in terms
of gauge-invariant projectors in the following way \cite{Mitscherling2025arXiv}:

\begin{equation}
\mathcal{Q}_{\mathbf{q}}^{ab}=\mathrm{Tr}\left(\hat{P}_{\mathbf{q}}\left(\partial^{a}\hat{P}_{\mathbf{q}}\right)\left(\partial^{b}\hat{P}_{\mathbf{q}}\right)\right)\,,\label{eq:Q_q^ab-alternative}
\end{equation}
where $\hat{P}_{\mathbf{q}}=\sum_{s\in\mathcal{S}}|u_{\mathbf{q}s}\rangle\langle u_{\mathbf{q}s}|$
is the projector onto the subspace of relevant Bloch bands.

Here we will use two well-established quasi-momentum space criteria
for $q$-ideal Chern bands \cite{Crepel2023PRR}:

(i) for all quasi-momenta in the Brillouin zone, the QGT has a constant \textcolor{black}{($\mathbf{q}$-independent)}
null vector \textcolor{black}{$w_\mathbf{q}\equiv w$}, i.e., $\mathcal{Q}_{\mathbf{q}}^{ab}w_{b}=0$, where
the summation over the repeated index $b$ is implied.

(ii) the integral $D_{{\rm QGT}}=\int_{{\rm 1BZ}}d^{2}\mathbf{q}\,\det\left(\mathcal{Q}_{\mathbf{q}}^{ab}\right)$
equals zero, i.e., $D_{{\rm QGT}}=0$.
% \db{Need to make q italic or not consistent}.

We find that for all quasi-momenta $\mathbf{q}$ in the Brillouin
zone, the quantum geometric tensor (QGT) possesses a null vector up
to numerical accuracy (eigenvalues deviate from zero by $\sim10^{-4}$).
For an ideal Chern band, the null vector $\overrightarrow{w}_{q}$
of QGT $\mathcal{\overleftrightarrow{Q}}_{\mathbf{q}}$ must be
$\mathbf{q}$-independent across the entire Brillouin zone. To
quantify the null vector's uniformity, we calculate its standard deviation
from the averaged value $\left\langle \overrightarrow{w}_{\mathbf{q}}\right\rangle $
given by: 

\begin{equation}
\sigma_{{\rm QGT}}=\frac{1}{A}\int_{{\rm 1BZ}}d^{2}\mathbf{q}\,\left|\overrightarrow{w}_{\mathbf{q}}-\left\langle \overrightarrow{w}_{\mathbf{q}}\right\rangle \right|^{2}\,,\label{eq:sigma_QGT}
\end{equation}
where  $A$ is the area of the first Brillouin zone. A small
value of the standard deviation, $\sigma_{{\rm QGT}}\ll1$, indicates
a nearly constant null vector.
\begin{figure}[tbh!]
\centering\includegraphics[width=1.0\columnwidth]{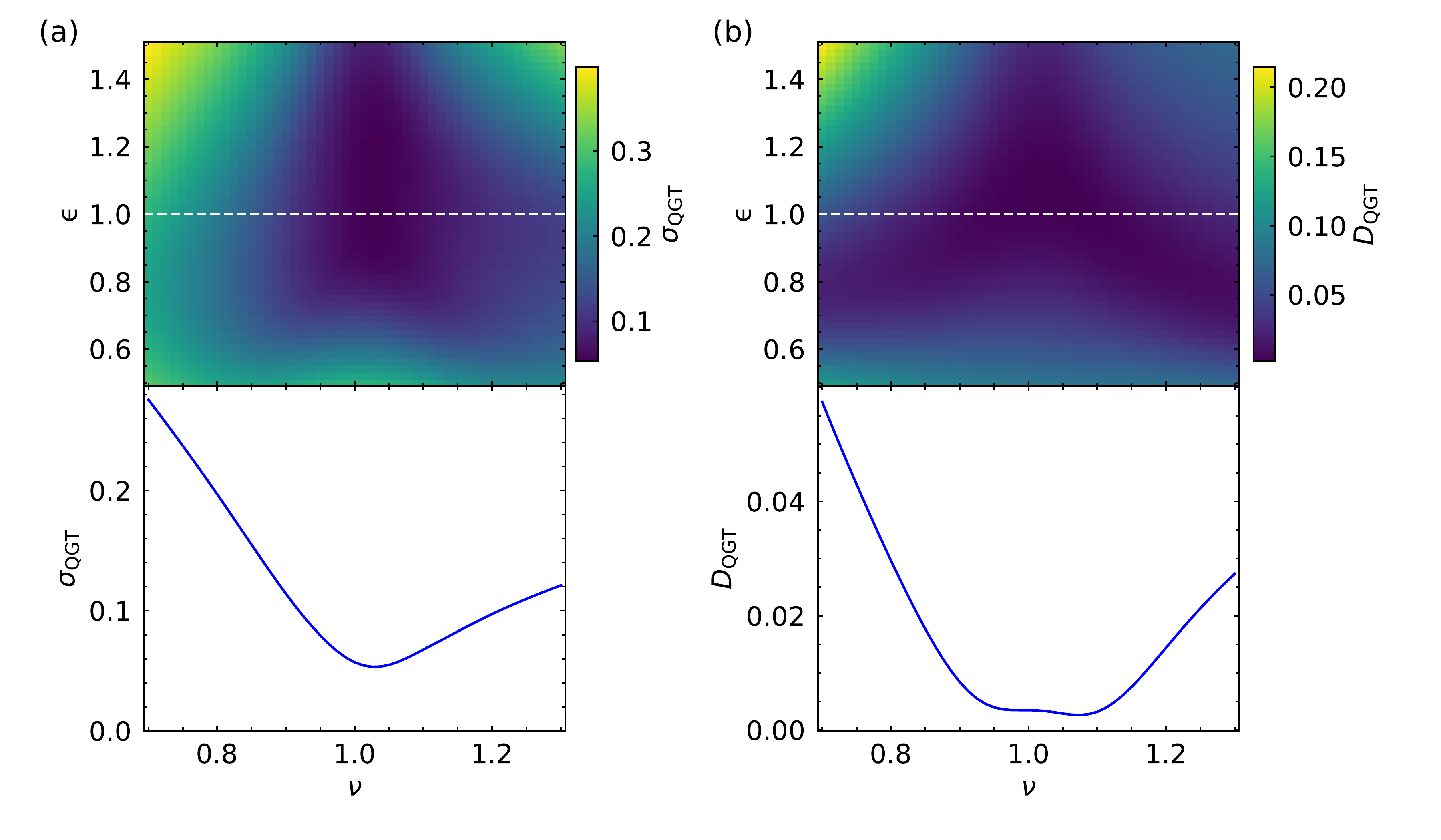}
% {\string"Article Figures/std_dev_psi_QGT_vs_c_epsilon_with_hline!2025-06-15!17.21.02\string".png}\includegraphics[width=0.5\columnwidth]{\string"Article Figures/D_QGT_vs_c_epsilon_with_hline!2025-06-15!17.21.02\string".png}
\caption{Dependence on $\epsilon$ and $\nu$ of the standard deviation of
the QGT null vector field $\sigma_{{\rm QGT}}$ of the lowest Bloch
band for $\Delta=2000E_{{\rm R}}$, $\Gamma=1000E_{{\rm R}}$, $\Omega_{0}=2000E_{{\rm R}}$.
}

\label{fig:std_dev_QGT-1}
\end{figure}

For parameter values near $\epsilon=\nu=1$, we observe
that the standard deviation $\sigma_{{\rm QGT}}$ is indeed much less than unity, it is below
0.1, as seen in Fig.~\ref{fig:std_dev_QGT-1}(a). Away from the aforementioned
parameter region, the null vector obtains a more significant $\mathbf{q}$-dependence
and the condition (i) for an ideal Chern band is no longer satisfied.
The condition (ii) also applies 
in a range of parameters around $\epsilon=\nu=1$,
as seen in Fig.~\ref{fig:std_dev_QGT-1}(b), but covers a somewhat
larger region of $\epsilon$-$\nu$ parameter space as compared to
Fig.~\ref{fig:std_dev_QGT-1}(a). It is noteworthy that in
Subsec.~\ref{subsec:Spectrum}, Fig.~\ref{fig:total-rho-c-grid}, the reorganization
of the ground state was observed around $\nu=0.95$ and $\epsilon=1$,
where a slight deviation from the optimal value of condition (i) is
seen in Fig.~\ref{fig:std_dev_QGT-1}(a).

\section{Concluding remarks\label{sec:Discussion}}

We have developed a general framework for engineering 2D sub-wavelength topological optical lattices for dark-state atoms. By properly designing the spatial profiles of the laser fields coupling atomic internal states in a $\Lambda$ scheme, one can generate Kronig-Penney-like subwavelength scalar potentials co-localised with intense, narrow synthetic magnetic flux tubes. The latter are compensated by a smooth background magnetic field of opposite sign, leading to zero net magnetic flux per unit cell, yet still enabling topologically nontrivial band structures.  Thus there are some similarities with the Haldane-type lattice model \cite{Haldane88PRL}
with zero net flux over an elementary cell, but non-trivial topology
due to non-zero fluxes over the individual plaquettes constituting
the elementary cell.  Returning to the present situation, for narrow enough flux peaks, the atoms effectively experience a uniform background magnetic field, resembling the Landau level problem. This supports nearly flat Bloch bands with unit Chern numbers, like in the quantum Hall systems.

We also verified that the system supports ideal Chern bands, as both Quantum Geometric Tensor conditions—null vector constancy and vanishing determinant integral—are satisfied near the perfect tuning point where $\epsilon$ and $\nu $ are close to unity. This makes the scheme well-suited for simulating the fractional Chern insulators in ultracold atomic gases, offering a platform for exploring strongly correlated topological phases with high tunability.

\section*{Acknowledgements}

G.J. thanks Jean Dalibard for helpful discussions on ref.~\cite{Dalibard24arXiv}.
The authors acknowledge support from the Lithuanian Research Council
(Grant No. S- MIP-24-97). This work has also received funding from
COST Action POLYTOPO CA23134, supported by COST (European Cooperation
in Science and Technology).

\newpage{}

\bibliography{2D_Dark_state_lattices}

\begin{thebibliography}{49}%
\makeatletter
\providecommand \@ifxundefined [1]{%
 \@ifx{#1\undefined}
}%
\providecommand \@ifnum [1]{%
 \ifnum #1\expandafter \@firstoftwo
 \else \expandafter \@secondoftwo
 \fi
}%
\providecommand \@ifx [1]{%
 \ifx #1\expandafter \@firstoftwo
 \else \expandafter \@secondoftwo
 \fi
}%
\providecommand \natexlab [1]{#1}%
\providecommand \enquote  [1]{``#1''}%
\providecommand \bibnamefont  [1]{#1}%
\providecommand \bibfnamefont [1]{#1}%
\providecommand \citenamefont [1]{#1}%
\providecommand \href@noop [0]{\@secondoftwo}%
\providecommand \href [0]{\begingroup \@sanitize@url \@href}%
\providecommand \@href[1]{\@@startlink{#1}\@@href}%
\providecommand \@@href[1]{\endgroup#1\@@endlink}%
\providecommand \@sanitize@url [0]{\catcode `\\12\catcode `\$12\catcode `\&12\catcode `\#12\catcode `\^12\catcode `\_12\catcode `\%12\relax}%
\providecommand \@@startlink[1]{}%
\providecommand \@@endlink[0]{}%
\providecommand \url  [0]{\begingroup\@sanitize@url \@url }%
\providecommand \@url [1]{\endgroup\@href {#1}{\urlprefix }}%
\providecommand \urlprefix  [0]{URL }%
\providecommand \Eprint [0]{\href }%
\providecommand \doibase [0]{https://doi.org/}%
\providecommand \selectlanguage [0]{\@gobble}%
\providecommand \bibinfo  [0]{\@secondoftwo}%
\providecommand \bibfield  [0]{\@secondoftwo}%
\providecommand \translation [1]{[#1]}%
\providecommand \BibitemOpen [0]{}%
\providecommand \bibitemStop [0]{}%
\providecommand \bibitemNoStop [0]{.\EOS\space}%
\providecommand \EOS [0]{\spacefactor3000\relax}%
\providecommand \BibitemShut  [1]{\csname bibitem#1\endcsname}%
\let\auto@bib@innerbib\@empty
%</preamble>
\bibitem [{\citenamefont {Qi}\ and\ \citenamefont {Zhang}(2011)}]{QiZhangRMP2011}%
  \BibitemOpen
  \bibfield  {author} {\bibinfo {author} {\bibfnamefont {X.-L.}\ \bibnamefont {Qi}}\ and\ \bibinfo {author} {\bibfnamefont {S.-C.}\ \bibnamefont {Zhang}},\ }\bibfield  {title} {\bibinfo {title} {Topological insulators and superconductors},\ }\href {https://doi.org/10.1103/RevModPhys.83.1057} {\bibfield  {journal} {\bibinfo  {journal} {Rev. Mod. Phys.}\ }\textbf {\bibinfo {volume} {83}},\ \bibinfo {pages} {1057} (\bibinfo {year} {2011})}\BibitemShut {NoStop}%
\bibitem [{\citenamefont {Hasan}\ and\ \citenamefont {Kane}(2010)}]{HasanKaneRMP2010}%
  \BibitemOpen
  \bibfield  {author} {\bibinfo {author} {\bibfnamefont {M.~Z.}\ \bibnamefont {Hasan}}\ and\ \bibinfo {author} {\bibfnamefont {C.~L.}\ \bibnamefont {Kane}},\ }\bibfield  {title} {\bibinfo {title} {\emph{Colloquium}: Topological insulators},\ }\href {https://doi.org/10.1103/RevModPhys.82.3045} {\bibfield  {journal} {\bibinfo  {journal} {Rev. Mod. Phys.}\ }\textbf {\bibinfo {volume} {82}},\ \bibinfo {pages} {3045} (\bibinfo {year} {2010})}\BibitemShut {NoStop}%
\bibitem [{\citenamefont {Dalibard}\ \emph {et~al.}(2011)\citenamefont {Dalibard}, \citenamefont {Gerbier}, \citenamefont {Juzeli\={u}nas},\ and\ \citenamefont {\"Ohberg}}]{Dalibard11RMP}%
  \BibitemOpen
  \bibfield  {author} {\bibinfo {author} {\bibfnamefont {J.}~\bibnamefont {Dalibard}}, \bibinfo {author} {\bibfnamefont {F.}~\bibnamefont {Gerbier}}, \bibinfo {author} {\bibfnamefont {G.}~\bibnamefont {Juzeli\={u}nas}},\ and\ \bibinfo {author} {\bibfnamefont {P.}~\bibnamefont {\"Ohberg}},\ }\bibfield  {title} {\bibinfo {title} {\emph{Colloquium}: Artificial gauge potentials for neutral atoms},\ }\href {https://doi.org/10.1103/RevModPhys.83.1523} {\bibfield  {journal} {\bibinfo  {journal} {Rev. Mod. Phys.}\ }\textbf {\bibinfo {volume} {83}},\ \bibinfo {pages} {1523} (\bibinfo {year} {2011})}\BibitemShut {NoStop}%
\bibitem [{\citenamefont {Goldman}\ \emph {et~al.}(2014)\citenamefont {Goldman}, \citenamefont {Juzeli{\={u}}nas}, \citenamefont {{\"O}hberg},\ and\ \citenamefont {Spielman}}]{Goldman2014}%
  \BibitemOpen
  \bibfield  {author} {\bibinfo {author} {\bibfnamefont {N.}~\bibnamefont {Goldman}}, \bibinfo {author} {\bibfnamefont {G.}~\bibnamefont {Juzeli{\={u}}nas}}, \bibinfo {author} {\bibfnamefont {P.}~\bibnamefont {{\"O}hberg}},\ and\ \bibinfo {author} {\bibfnamefont {I.~B.}\ \bibnamefont {Spielman}},\ }\bibfield  {title} {\bibinfo {title} {Light-induced gauge fields for ultracold atoms},\ }\href {https://doi.org/10.1088/0034-4885/77/12/126401} {\bibfield  {journal} {\bibinfo  {journal} {Rep. Prog. Phys.}\ }\textbf {\bibinfo {volume} {77}},\ \bibinfo {pages} {126401} (\bibinfo {year} {2014})}\BibitemShut {NoStop}%
\bibitem [{\citenamefont {Aidelsburger}\ \emph {et~al.}(2018)\citenamefont {Aidelsburger}, \citenamefont {Nascimbene},\ and\ \citenamefont {Goldman}}]{Aidelsburger2018Physique}%
  \BibitemOpen
  \bibfield  {author} {\bibinfo {author} {\bibfnamefont {M.}~\bibnamefont {Aidelsburger}}, \bibinfo {author} {\bibfnamefont {S.}~\bibnamefont {Nascimbene}},\ and\ \bibinfo {author} {\bibfnamefont {N.}~\bibnamefont {Goldman}},\ }\bibfield  {title} {\bibinfo {title} {Artificial gauge fields in materials and engineered systems},\ }\href {https://doi.org/https://doi.org/10.1016/j.crhy.2018.03.002} {\bibfield  {journal} {\bibinfo  {journal} {Comptes Rendus Physique}\ }\textbf {\bibinfo {volume} {19}},\ \bibinfo {pages} {394} (\bibinfo {year} {2018})},\ \bibinfo {note} {quantum simulation / Simulation quantique}\BibitemShut {NoStop}%
\bibitem [{\citenamefont {Galitski}\ \emph {et~al.}(2019)\citenamefont {Galitski}, \citenamefont {Juzeli{\=u}nas},\ and\ \citenamefont {Spielman}}]{Galitski19PT}%
  \BibitemOpen
  \bibfield  {author} {\bibinfo {author} {\bibfnamefont {V.}~\bibnamefont {Galitski}}, \bibinfo {author} {\bibfnamefont {G.}~\bibnamefont {Juzeli{\=u}nas}},\ and\ \bibinfo {author} {\bibfnamefont {I.~B.}\ \bibnamefont {Spielman}},\ }\bibfield  {title} {\bibinfo {title} {Artificial gauge fields with ultracold atoms},\ }\href@noop {} {\bibfield  {journal} {\bibinfo  {journal} {Phys. Today}\ }\textbf {\bibinfo {volume} {72}},\ \bibinfo {pages} {38} (\bibinfo {year} {2019})}\BibitemShut {NoStop}%
\bibitem [{\citenamefont {Cooper}\ \emph {et~al.}(2019)\citenamefont {Cooper}, \citenamefont {Dalibard},\ and\ \citenamefont {Spielman}}]{Cooper2019RMP}%
  \BibitemOpen
  \bibfield  {author} {\bibinfo {author} {\bibfnamefont {N.~R.}\ \bibnamefont {Cooper}}, \bibinfo {author} {\bibfnamefont {J.}~\bibnamefont {Dalibard}},\ and\ \bibinfo {author} {\bibfnamefont {I.~B.}\ \bibnamefont {Spielman}},\ }\bibfield  {title} {\bibinfo {title} {Topological bands for ultracold atoms},\ }\href {https://doi.org/10.1103/RevModPhys.91.015005} {\bibfield  {journal} {\bibinfo  {journal} {Rev. Mod. Phys.}\ }\textbf {\bibinfo {volume} {91}},\ \bibinfo {pages} {015005} (\bibinfo {year} {2019})}\BibitemShut {NoStop}%
\bibitem [{\citenamefont {Ozawa}\ \emph {et~al.}(2019)\citenamefont {Ozawa}, \citenamefont {Price}, \citenamefont {Amo}, \citenamefont {Goldman}, \citenamefont {Hafezi}, \citenamefont {Lu}, \citenamefont {Rechtsman}, \citenamefont {Schuster}, \citenamefont {Simon}, \citenamefont {Zilberberg},\ and\ \citenamefont {Carusotto}}]{Ozawa2019RMP}%
  \BibitemOpen
  \bibfield  {author} {\bibinfo {author} {\bibfnamefont {T.}~\bibnamefont {Ozawa}}, \bibinfo {author} {\bibfnamefont {H.~M.}\ \bibnamefont {Price}}, \bibinfo {author} {\bibfnamefont {A.}~\bibnamefont {Amo}}, \bibinfo {author} {\bibfnamefont {N.}~\bibnamefont {Goldman}}, \bibinfo {author} {\bibfnamefont {M.}~\bibnamefont {Hafezi}}, \bibinfo {author} {\bibfnamefont {L.}~\bibnamefont {Lu}}, \bibinfo {author} {\bibfnamefont {M.~C.}\ \bibnamefont {Rechtsman}}, \bibinfo {author} {\bibfnamefont {D.}~\bibnamefont {Schuster}}, \bibinfo {author} {\bibfnamefont {J.}~\bibnamefont {Simon}}, \bibinfo {author} {\bibfnamefont {O.}~\bibnamefont {Zilberberg}},\ and\ \bibinfo {author} {\bibfnamefont {I.}~\bibnamefont {Carusotto}},\ }\bibfield  {title} {\bibinfo {title} {Topological photonics},\ }\href {https://doi.org/10.1103/RevModPhys.91.015006} {\bibfield  {journal} {\bibinfo  {journal} {Rev. Mod. Phys.}\ }\textbf {\bibinfo {volume} {91}},\ \bibinfo {pages} {015006} (\bibinfo {year} {2019})}\BibitemShut {NoStop}%
\bibitem [{\citenamefont {Shen}\ \emph {et~al.}(2024)\citenamefont {Shen}, \citenamefont {Zhang}, \citenamefont {Shi}, \citenamefont {Du}, \citenamefont {Yuan},\ and\ \citenamefont {Zayats}}]{Shen2024}%
  \BibitemOpen
  \bibfield  {author} {\bibinfo {author} {\bibfnamefont {Y.}~\bibnamefont {Shen}}, \bibinfo {author} {\bibfnamefont {Q.}~\bibnamefont {Zhang}}, \bibinfo {author} {\bibfnamefont {P.}~\bibnamefont {Shi}}, \bibinfo {author} {\bibfnamefont {L.}~\bibnamefont {Du}}, \bibinfo {author} {\bibfnamefont {X.}~\bibnamefont {Yuan}},\ and\ \bibinfo {author} {\bibfnamefont {A.~V.}\ \bibnamefont {Zayats}},\ }\bibfield  {title} {\bibinfo {title} {Optical skyrmions and other topological quasiparticles of light},\ }\href {https://doi.org/10.1038/s41566-023-01325-7} {\bibfield  {journal} {\bibinfo  {journal} {Nature Photonics}\ }\textbf {\bibinfo {volume} {18}},\ \bibinfo {pages} {15} (\bibinfo {year} {2024})}\BibitemShut {NoStop}%
\bibitem [{\citenamefont {McWilliam}\ \emph {et~al.}(2023)\citenamefont {McWilliam}, \citenamefont {Cisowski}, \citenamefont {Ye}, \citenamefont {Speirits}, \citenamefont {Götte}, \citenamefont {Barnett},\ and\ \citenamefont {Franke-Arnold}}]{Sonja2023}%
  \BibitemOpen
  \bibfield  {author} {\bibinfo {author} {\bibfnamefont {A.}~\bibnamefont {McWilliam}}, \bibinfo {author} {\bibfnamefont {C.~M.}\ \bibnamefont {Cisowski}}, \bibinfo {author} {\bibfnamefont {Z.}~\bibnamefont {Ye}}, \bibinfo {author} {\bibfnamefont {F.~C.}\ \bibnamefont {Speirits}}, \bibinfo {author} {\bibfnamefont {J.~B.}\ \bibnamefont {Götte}}, \bibinfo {author} {\bibfnamefont {S.~M.}\ \bibnamefont {Barnett}},\ and\ \bibinfo {author} {\bibfnamefont {S.}~\bibnamefont {Franke-Arnold}},\ }\bibfield  {title} {\bibinfo {title} {Topological approach of characterizing optical skyrmions and multi-skyrmions},\ }\href {https://doi.org/https://doi.org/10.1002/lpor.202300155} {\bibfield  {journal} {\bibinfo  {journal} {Laser Photonics Rev.}\ }\textbf {\bibinfo {volume} {17}},\ \bibinfo {pages} {2300155} (\bibinfo {year} {2023})}\BibitemShut {NoStop}%
\bibitem [{\citenamefont {Xue}\ \emph {et~al.}(2022)\citenamefont {Xue}, \citenamefont {Yang},\ and\ \citenamefont {Zhang}}]{Xue2022NatRev}%
  \BibitemOpen
  \bibfield  {author} {\bibinfo {author} {\bibfnamefont {H.}~\bibnamefont {Xue}}, \bibinfo {author} {\bibfnamefont {Y.}~\bibnamefont {Yang}},\ and\ \bibinfo {author} {\bibfnamefont {B.}~\bibnamefont {Zhang}},\ }\bibfield  {title} {\bibinfo {title} {Topological acoustics},\ }\href {https://doi.org/10.1038/s41578-022-00465-6} {\bibfield  {journal} {\bibinfo  {journal} {Nature Reviews Materials}\ }\textbf {\bibinfo {volume} {7}},\ \bibinfo {pages} {974} (\bibinfo {year} {2022})}\BibitemShut {NoStop}%
\bibitem [{\citenamefont {Guo}\ \emph {et~al.}(2025)\citenamefont {Guo}, \citenamefont {Jezequel}, \citenamefont {Padlewski}, \citenamefont {Lissek}, \citenamefont {Delplace},\ and\ \citenamefont {Fleury}}]{XinxinGuo2025}%
  \BibitemOpen
  \bibfield  {author} {\bibinfo {author} {\bibfnamefont {X.}~\bibnamefont {Guo}}, \bibinfo {author} {\bibfnamefont {L.}~\bibnamefont {Jezequel}}, \bibinfo {author} {\bibfnamefont {M.}~\bibnamefont {Padlewski}}, \bibinfo {author} {\bibfnamefont {H.}~\bibnamefont {Lissek}}, \bibinfo {author} {\bibfnamefont {P.}~\bibnamefont {Delplace}},\ and\ \bibinfo {author} {\bibfnamefont {R.}~\bibnamefont {Fleury}},\ }\bibfield  {title} {\bibinfo {title} {{Practical realization of chiral nonlinearity for strong topological protection}},\ }\href {https://doi.org/10.21468/SciPostPhys.18.1.034} {\bibfield  {journal} {\bibinfo  {journal} {SciPost Phys.}\ }\textbf {\bibinfo {volume} {18}},\ \bibinfo {pages} {034} (\bibinfo {year} {2025})}\BibitemShut {NoStop}%
\bibitem [{\citenamefont {Lewenstein}\ \emph {et~al.}(2007)\citenamefont {Lewenstein}, \citenamefont {Sanpera}, \citenamefont {Ahufinger}, \citenamefont {Damski}, \citenamefont {Sen(De)},\ and\ \citenamefont {Sen}}]{Lewenstein2007}%
  \BibitemOpen
  \bibfield  {author} {\bibinfo {author} {\bibfnamefont {M.}~\bibnamefont {Lewenstein}}, \bibinfo {author} {\bibfnamefont {A.}~\bibnamefont {Sanpera}}, \bibinfo {author} {\bibfnamefont {V.}~\bibnamefont {Ahufinger}}, \bibinfo {author} {\bibfnamefont {B.}~\bibnamefont {Damski}}, \bibinfo {author} {\bibfnamefont {A.}~\bibnamefont {Sen(De)}},\ and\ \bibinfo {author} {\bibfnamefont {U.}~\bibnamefont {Sen}},\ }\bibfield  {title} {\bibinfo {title} {Ultracold atomic gases in optical lattices: {M}imicking condensed matter physics and beyond},\ }\bibfield  {booktitle} {\emph {\bibinfo {booktitle} {Advances in Physics}},\ }\href {https://doi.org/10.1080/00018730701223200} {\bibfield  {journal} {\bibinfo  {journal} {Adv. Phys.}\ }\textbf {\bibinfo {volume} {56}},\ \bibinfo {pages} {243} (\bibinfo {year} {2007})}\BibitemShut {NoStop}%
\bibitem [{\citenamefont {Bloch}\ \emph {et~al.}(2008)\citenamefont {Bloch}, \citenamefont {Dalibard},\ and\ \citenamefont {Zwerger}}]{Bloch2008}%
  \BibitemOpen
  \bibfield  {author} {\bibinfo {author} {\bibfnamefont {I.}~\bibnamefont {Bloch}}, \bibinfo {author} {\bibfnamefont {J.}~\bibnamefont {Dalibard}},\ and\ \bibinfo {author} {\bibfnamefont {W.}~\bibnamefont {Zwerger}},\ }\bibfield  {title} {\bibinfo {title} {Many-body physics with ultracold gases},\ }\href@noop {} {\bibfield  {journal} {\bibinfo  {journal} {Rev. Mod. Phys.}\ }\textbf {\bibinfo {volume} {80}},\ \bibinfo {pages} {885} (\bibinfo {year} {2008})}\BibitemShut {NoStop}%
\bibitem [{\citenamefont {Dutta}\ \emph {et~al.}(2015)\citenamefont {Dutta}, \citenamefont {Gajda}, \citenamefont {Hauke}, \citenamefont {Lewenstein}, \citenamefont {L{\"u}hmann}, \citenamefont {Malomed}, \citenamefont {Sowi{\'n}ski},\ and\ \citenamefont {Zakrzewski}}]{Dutta15RoPP}%
  \BibitemOpen
  \bibfield  {author} {\bibinfo {author} {\bibfnamefont {O.}~\bibnamefont {Dutta}}, \bibinfo {author} {\bibfnamefont {M.}~\bibnamefont {Gajda}}, \bibinfo {author} {\bibfnamefont {P.}~\bibnamefont {Hauke}}, \bibinfo {author} {\bibfnamefont {M.}~\bibnamefont {Lewenstein}}, \bibinfo {author} {\bibfnamefont {D.-S.}\ \bibnamefont {L{\"u}hmann}}, \bibinfo {author} {\bibfnamefont {B.~A.}\ \bibnamefont {Malomed}}, \bibinfo {author} {\bibfnamefont {T.}~\bibnamefont {Sowi{\'n}ski}},\ and\ \bibinfo {author} {\bibfnamefont {J.}~\bibnamefont {Zakrzewski}},\ }\bibfield  {title} {\bibinfo {title} {Non-standard {H}ubbard models in optical lattices: {A} review},\ }\href {https://doi.org/10.1088/0034-4885/78/6/066001} {\bibfield  {journal} {\bibinfo  {journal} {Reports on Progress in Physics}\ }\textbf {\bibinfo {volume} {78}},\ \bibinfo {pages} {066001} (\bibinfo {year} {2015})}\BibitemShut {NoStop}%
\bibitem [{\citenamefont {Gross}\ and\ \citenamefont {Bloch}(2017)}]{Gross-Bloch17Science}%
  \BibitemOpen
  \bibfield  {author} {\bibinfo {author} {\bibfnamefont {C.}~\bibnamefont {Gross}}\ and\ \bibinfo {author} {\bibfnamefont {I.}~\bibnamefont {Bloch}},\ }\bibfield  {title} {\bibinfo {title} {Quantum simulations with ultracold atoms in optical lattices},\ }\href {https://doi.org/10.1126/science.aal3837} {\bibfield  {journal} {\bibinfo  {journal} {Science}\ }\textbf {\bibinfo {volume} {357}},\ \bibinfo {pages} {995} (\bibinfo {year} {2017})}\BibitemShut {NoStop}%
\bibitem [{\citenamefont {Sch{\"a}fer}\ \emph {et~al.}(2020)\citenamefont {Sch{\"a}fer}, \citenamefont {Fukuhara}, \citenamefont {Sugawa}, \citenamefont {Takasu},\ and\ \citenamefont {Takahashi}}]{Takahashi20NatRev}%
  \BibitemOpen
  \bibfield  {author} {\bibinfo {author} {\bibfnamefont {F.}~\bibnamefont {Sch{\"a}fer}}, \bibinfo {author} {\bibfnamefont {T.}~\bibnamefont {Fukuhara}}, \bibinfo {author} {\bibfnamefont {S.}~\bibnamefont {Sugawa}}, \bibinfo {author} {\bibfnamefont {Y.}~\bibnamefont {Takasu}},\ and\ \bibinfo {author} {\bibfnamefont {Y.}~\bibnamefont {Takahashi}},\ }\bibfield  {title} {\bibinfo {title} {Tools for quantum simulation with ultracold atoms in optical lattices},\ }\href {https://doi.org/10.1038/s42254-020-0195-3} {\bibfield  {journal} {\bibinfo  {journal} {Nature Reviews Physics}\ }\textbf {\bibinfo {volume} {2}},\ \bibinfo {pages} {411} (\bibinfo {year} {2020})}\BibitemShut {NoStop}%
\bibitem [{\citenamefont {Scully}\ and\ \citenamefont {Zubairy}(1997)}]{Scully2008}%
  \BibitemOpen
  \bibfield  {author} {\bibinfo {author} {\bibfnamefont {M.~O.}\ \bibnamefont {Scully}}\ and\ \bibinfo {author} {\bibfnamefont {M.~S.}\ \bibnamefont {Zubairy}},\ }\href@noop {} {\emph {\bibinfo {title} {Quantum {O}ptics}}}\ (\bibinfo  {publisher} {Cambridge University Press, Cambridge},\ \bibinfo {year} {1997})\BibitemShut {NoStop}%
\bibitem [{\citenamefont {Vitanov}\ \emph {et~al.}(2017)\citenamefont {Vitanov}, \citenamefont {Rangelov}, \citenamefont {Shore},\ and\ \citenamefont {Bergmann}}]{Bergman2017}%
  \BibitemOpen
  \bibfield  {author} {\bibinfo {author} {\bibfnamefont {N.~V.}\ \bibnamefont {Vitanov}}, \bibinfo {author} {\bibfnamefont {A.~A.}\ \bibnamefont {Rangelov}}, \bibinfo {author} {\bibfnamefont {B.~W.}\ \bibnamefont {Shore}},\ and\ \bibinfo {author} {\bibfnamefont {K.}~\bibnamefont {Bergmann}},\ }\bibfield  {title} {\bibinfo {title} {Stimulated raman adiabatic passage in physics, chemistry, and beyond},\ }\href {https://doi.org/10.1103/RevModPhys.89.015006} {\bibfield  {journal} {\bibinfo  {journal} {Rev. Mod. Phys.}\ }\textbf {\bibinfo {volume} {89}},\ \bibinfo {pages} {015006} (\bibinfo {year} {2017})}\BibitemShut {NoStop}%
\bibitem [{\citenamefont {\L{}\k{a}cki}\ \emph {et~al.}(2016)\citenamefont {\L{}\k{a}cki}, \citenamefont {Baranov}, \citenamefont {Pichler},\ and\ \citenamefont {Zoller}}]{Zoller2016}%
  \BibitemOpen
  \bibfield  {author} {\bibinfo {author} {\bibfnamefont {M.}~\bibnamefont {\L{}\k{a}cki}}, \bibinfo {author} {\bibfnamefont {M.~A.}\ \bibnamefont {Baranov}}, \bibinfo {author} {\bibfnamefont {H.}~\bibnamefont {Pichler}},\ and\ \bibinfo {author} {\bibfnamefont {P.}~\bibnamefont {Zoller}},\ }\bibfield  {title} {\bibinfo {title} {Nanoscale ''dark state'' optical potentials for cold atoms},\ }\href {https://doi.org/10.1103/physrevlett.117.233001} {\bibfield  {journal} {\bibinfo  {journal} {Phys. Rev. Lett.}\ }\textbf {\bibinfo {volume} {117}},\ \bibinfo {pages} {233001} (\bibinfo {year} {2016})}\BibitemShut {NoStop}%
\bibitem [{\citenamefont {Jendrzejewski}\ \emph {et~al.}(2016)\citenamefont {Jendrzejewski}, \citenamefont {Eckel}, \citenamefont {Tiecke}, \citenamefont {Juzeli\={u}nas}, \citenamefont {Campbell}, \citenamefont {Jiang},\ and\ \citenamefont {Gorshkov}}]{Jendrzejewski2016}%
  \BibitemOpen
  \bibfield  {author} {\bibinfo {author} {\bibfnamefont {F.}~\bibnamefont {Jendrzejewski}}, \bibinfo {author} {\bibfnamefont {S.}~\bibnamefont {Eckel}}, \bibinfo {author} {\bibfnamefont {T.~G.}\ \bibnamefont {Tiecke}}, \bibinfo {author} {\bibfnamefont {G.}~\bibnamefont {Juzeli\={u}nas}}, \bibinfo {author} {\bibfnamefont {G.~K.}\ \bibnamefont {Campbell}}, \bibinfo {author} {\bibfnamefont {L.}~\bibnamefont {Jiang}},\ and\ \bibinfo {author} {\bibfnamefont {A.~V.}\ \bibnamefont {Gorshkov}},\ }\bibfield  {title} {\bibinfo {title} {Subwavelength-width optical tunnel junctions for ultracold atoms},\ }\href {https://doi.org/10.1103/physreva.94.063422} {\bibfield  {journal} {\bibinfo  {journal} {Phys. Rev. A}\ }\textbf {\bibinfo {volume} {94}},\ \bibinfo {pages} {063422} (\bibinfo {year} {2016})}\BibitemShut {NoStop}%
\bibitem [{\citenamefont {Ge}\ and\ \citenamefont {Zubairy}(2020)}]{Zubairy2020}%
  \BibitemOpen
  \bibfield  {author} {\bibinfo {author} {\bibfnamefont {W.}~\bibnamefont {Ge}}\ and\ \bibinfo {author} {\bibfnamefont {M.~S.}\ \bibnamefont {Zubairy}},\ }\bibfield  {title} {\bibinfo {title} {Dark-state optical potential barriers with nanoscale spacing},\ }\href {https://doi.org/10.1103/PhysRevA.101.023403} {\bibfield  {journal} {\bibinfo  {journal} {Phys. Rev. A}\ }\textbf {\bibinfo {volume} {101}},\ \bibinfo {pages} {023403} (\bibinfo {year} {2020})}\BibitemShut {NoStop}%
\bibitem [{\citenamefont {Kubala}\ \emph {et~al.}(2021)\citenamefont {Kubala}, \citenamefont {Zakrzewski},\ and\ \citenamefont {\L{}\k{a}cki}}]{Kubala2021}%
  \BibitemOpen
  \bibfield  {author} {\bibinfo {author} {\bibfnamefont {P.}~\bibnamefont {Kubala}}, \bibinfo {author} {\bibfnamefont {J.}~\bibnamefont {Zakrzewski}},\ and\ \bibinfo {author} {\bibfnamefont {M.}~\bibnamefont {\L{}\k{a}cki}},\ }\bibfield  {title} {\bibinfo {title} {Optical lattice for a tripodlike atomic level structure},\ }\href {https://doi.org/10.1103/PhysRevA.104.053312} {\bibfield  {journal} {\bibinfo  {journal} {Phys. Rev. A}\ }\textbf {\bibinfo {volume} {104}},\ \bibinfo {pages} {053312} (\bibinfo {year} {2021})}\BibitemShut {NoStop}%
\bibitem [{\citenamefont {Gvozdiovas}\ \emph {et~al.}(2021)\citenamefont {Gvozdiovas}, \citenamefont {Ra{\v c}kauskas},\ and\ \citenamefont {Juzeli{\=u}nas}}]{Gvozdiovas2021}%
  \BibitemOpen
  \bibfield  {author} {\bibinfo {author} {\bibfnamefont {E.}~\bibnamefont {Gvozdiovas}}, \bibinfo {author} {\bibfnamefont {P.}~\bibnamefont {Ra{\v c}kauskas}},\ and\ \bibinfo {author} {\bibfnamefont {G.}~\bibnamefont {Juzeli{\=u}nas}},\ }\bibfield  {title} {\bibinfo {title} {{Optical lattice with spin-dependent sub-wavelength barriers}},\ }\href {https://doi.org/10.21468/SciPostPhys.11.6.100} {\bibfield  {journal} {\bibinfo  {journal} {SciPost Phys.}\ }\textbf {\bibinfo {volume} {11}},\ \bibinfo {pages} {100} (\bibinfo {year} {2021})}\BibitemShut {NoStop}%
\bibitem [{\citenamefont {Wang}\ \emph {et~al.}(2018)\citenamefont {Wang}, \citenamefont {Subhankar}, \citenamefont {Bienias}, \citenamefont {\L{}\k{a}cki}, \citenamefont {Tsui}, \citenamefont {Baranov}, \citenamefont {Gorshkov}, \citenamefont {Zoller}, \citenamefont {Porto},\ and\ \citenamefont {Rolston}}]{Wang2018}%
  \BibitemOpen
  \bibfield  {author} {\bibinfo {author} {\bibfnamefont {Y.}~\bibnamefont {Wang}}, \bibinfo {author} {\bibfnamefont {S.}~\bibnamefont {Subhankar}}, \bibinfo {author} {\bibfnamefont {P.}~\bibnamefont {Bienias}}, \bibinfo {author} {\bibfnamefont {M.}~\bibnamefont {\L{}\k{a}cki}}, \bibinfo {author} {\bibfnamefont {T.-C.}\ \bibnamefont {Tsui}}, \bibinfo {author} {\bibfnamefont {M.~A.}\ \bibnamefont {Baranov}}, \bibinfo {author} {\bibfnamefont {A.~V.}\ \bibnamefont {Gorshkov}}, \bibinfo {author} {\bibfnamefont {P.}~\bibnamefont {Zoller}}, \bibinfo {author} {\bibfnamefont {J.~V.}\ \bibnamefont {Porto}},\ and\ \bibinfo {author} {\bibfnamefont {S.~L.}\ \bibnamefont {Rolston}},\ }\bibfield  {title} {\bibinfo {title} {Dark state optical lattice with a subwavelength spatial structure},\ }\href {https://doi.org/10.1103/PhysRevLett.120.083601} {\bibfield  {journal} {\bibinfo  {journal} {Phys. Rev. Lett.}\ }\textbf {\bibinfo {volume} {120}},\ \bibinfo {pages} {083601} (\bibinfo {year} {2018})}\BibitemShut {NoStop}%
\bibitem [{\citenamefont {Tsui}\ \emph {et~al.}(2020)\citenamefont {Tsui}, \citenamefont {Wang}, \citenamefont {Subhankar}, \citenamefont {Porto},\ and\ \citenamefont {Rolston}}]{Tsui2020}%
  \BibitemOpen
  \bibfield  {author} {\bibinfo {author} {\bibfnamefont {T.-C.}\ \bibnamefont {Tsui}}, \bibinfo {author} {\bibfnamefont {Y.}~\bibnamefont {Wang}}, \bibinfo {author} {\bibfnamefont {S.}~\bibnamefont {Subhankar}}, \bibinfo {author} {\bibfnamefont {J.~V.}\ \bibnamefont {Porto}},\ and\ \bibinfo {author} {\bibfnamefont {S.~L.}\ \bibnamefont {Rolston}},\ }\bibfield  {title} {\bibinfo {title} {Realization of a stroboscopic optical lattice for cold atoms with subwavelength spacing},\ }\href {https://doi.org/10.1103/PhysRevA.101.041603} {\bibfield  {journal} {\bibinfo  {journal} {Phys. Rev. A}\ }\textbf {\bibinfo {volume} {101}},\ \bibinfo {pages} {041603(R)} (\bibinfo {year} {2020})}\BibitemShut {NoStop}%
\bibitem [{\citenamefont {Gvozdiovas}\ \emph {et~al.}(2023)\citenamefont {Gvozdiovas}, \citenamefont {Spielman},\ and\ \citenamefont {Juzeli\={u}nas}}]{Gvozdiovas23PRA}%
  \BibitemOpen
  \bibfield  {author} {\bibinfo {author} {\bibfnamefont {E.}~\bibnamefont {Gvozdiovas}}, \bibinfo {author} {\bibfnamefont {I.~B.}\ \bibnamefont {Spielman}},\ and\ \bibinfo {author} {\bibfnamefont {G.}~\bibnamefont {Juzeli\={u}nas}},\ }\bibfield  {title} {\bibinfo {title} {Interference-induced anisotropy in a two-dimensional dark-state optical lattice},\ }\href {https://doi.org/10.1103/PhysRevA.107.033328} {\bibfield  {journal} {\bibinfo  {journal} {Phys. Rev. A}\ }\textbf {\bibinfo {volume} {107}},\ \bibinfo {pages} {033328} (\bibinfo {year} {2023})}\BibitemShut {NoStop}%
\bibitem [{\citenamefont {Nascimbene}\ and\ \citenamefont {Dalibard}(2024)}]{Dalibard24arXiv}%
  \BibitemOpen
  \bibfield  {author} {\bibinfo {author} {\bibfnamefont {S.}~\bibnamefont {Nascimbene}}\ and\ \bibinfo {author} {\bibfnamefont {J.}~\bibnamefont {Dalibard}},\ }\href {https://arxiv.org/abs/2412.15038} {\bibinfo {title} {Emergence of a {L}andau level structure in dark optical lattices}} (\bibinfo {year} {2024}),\ \Eprint {https://arxiv.org/abs/2412.15038} {arXiv:2412.15038 [cond-mat.quant-gas]} \BibitemShut {NoStop}%
\bibitem [{\citenamefont {Sommer}\ and\ \citenamefont {Cooper}(2025)}]{Cooper2025arXiv}%
  \BibitemOpen
  \bibfield  {author} {\bibinfo {author} {\bibfnamefont {O.~E.}\ \bibnamefont {Sommer}}\ and\ \bibinfo {author} {\bibfnamefont {N.~R.}\ \bibnamefont {Cooper}},\ }\href {https://arxiv.org/abs/2509.01481} {\bibinfo {title} {Ideal optical flux lattices}} (\bibinfo {year} {2025}),\ \Eprint {https://arxiv.org/abs/2509.01481} {arXiv:2509.01481 [cond-mat.quant-gas]} \BibitemShut {NoStop}%
\bibitem [{\citenamefont {Landau}(1930)}]{Landau1930}%
  \BibitemOpen
  \bibfield  {author} {\bibinfo {author} {\bibfnamefont {L.}~\bibnamefont {Landau}},\ }\bibfield  {title} {\bibinfo {title} {Diamagnetismus der {M}etalle},\ }\href {https://doi.org/10.1007/BF01397213} {\bibfield  {journal} {\bibinfo  {journal} {Zeitschrift f{\"u}r Physik}\ }\textbf {\bibinfo {volume} {64}},\ \bibinfo {pages} {629} (\bibinfo {year} {1930})}\BibitemShut {NoStop}%
\bibitem [{\citenamefont {Landau}\ and\ \citenamefont {Lifshitz}(1977)}]{landau1977quantum}%
  \BibitemOpen
  \bibfield  {author} {\bibinfo {author} {\bibfnamefont {L.~D.}\ \bibnamefont {Landau}}\ and\ \bibinfo {author} {\bibfnamefont {E.~M.}\ \bibnamefont {Lifshitz}},\ }\href {https://doi.org/10.1016/C2013-0-02793-4} {\emph {\bibinfo {title} {Quantum {M}echanics: {N}on-{R}elativistic {T}heory}}},\ \bibinfo {edition} {3rd}\ ed.\ (\bibinfo  {publisher} {Pergamon},\ \bibinfo {address} {Oxford},\ \bibinfo {year} {1977})\BibitemShut {NoStop}%
\bibitem [{\citenamefont {Hern\'andez~Yanes}\ \emph {et~al.}(2022)\citenamefont {Hern\'andez~Yanes}, \citenamefont {P\l{}odzie\'{n}}, \citenamefont {Mackoit~Sinkevi\v{c}ien\.{e}}, \citenamefont {\v{Z}labys}, \citenamefont {Juzeli\={u}nas},\ and\ \citenamefont {Witkowska}}]{Yanes22PRL}%
  \BibitemOpen
  \bibfield  {author} {\bibinfo {author} {\bibfnamefont {T.}~\bibnamefont {Hern\'andez~Yanes}}, \bibinfo {author} {\bibfnamefont {M.}~\bibnamefont {P\l{}odzie\'{n}}}, \bibinfo {author} {\bibfnamefont {M.}~\bibnamefont {Mackoit~Sinkevi\v{c}ien\.{e}}}, \bibinfo {author} {\bibfnamefont {G.}~\bibnamefont {\v{Z}labys}}, \bibinfo {author} {\bibfnamefont {G.}~\bibnamefont {Juzeli\={u}nas}},\ and\ \bibinfo {author} {\bibfnamefont {E.}~\bibnamefont {Witkowska}},\ }\bibfield  {title} {\bibinfo {title} {One- and two-axis squeezing via laser coupling in an atomic {F}ermi-{H}ubbard model},\ }\href {https://doi.org/10.1103/PhysRevLett.129.090403} {\bibfield  {journal} {\bibinfo  {journal} {Phys. Rev. Lett.}\ }\textbf {\bibinfo {volume} {129}},\ \bibinfo {pages} {090403} (\bibinfo {year} {2022})}\BibitemShut {NoStop}%
\bibitem [{\citenamefont {Dum}\ and\ \citenamefont {Olshanii}(1996)}]{Dum96PRL}%
  \BibitemOpen
  \bibfield  {author} {\bibinfo {author} {\bibfnamefont {R.}~\bibnamefont {Dum}}\ and\ \bibinfo {author} {\bibfnamefont {M.}~\bibnamefont {Olshanii}},\ }\bibfield  {title} {\bibinfo {title} {Gauge structures in atom-laser interaction: {B}loch oscillations in a dark lattice},\ }\href {https://doi.org/10.1103/PhysRevLett.76.1788} {\bibfield  {journal} {\bibinfo  {journal} {Phys. Rev. Lett.}\ }\textbf {\bibinfo {volume} {76}},\ \bibinfo {pages} {1788} (\bibinfo {year} {1996})}\BibitemShut {NoStop}%
\bibitem [{Note1()}]{Note1}%
  \BibitemOpen
  \bibinfo {note} {This applies to the arXiv version of ref.~\cite {Dalibard24arXiv}, which was available during the original submission of our manuscript to Physical Review Research and to arXiv. Subsequently, a lossy excited level of the Lambda scheme was added in the “End Matter” of the revised ref.~\cite {Dalibard24arXiv}.}\BibitemShut {Stop}%
\bibitem [{\citenamefont {Juzeli\ifmmode~\bar{u}\else \={u}\fi{}nas}\ \emph {et~al.}(2005)\citenamefont {Juzeli\ifmmode~\bar{u}\else \={u}\fi{}nas}, \citenamefont {\"Ohberg}, \citenamefont {Ruseckas},\ and\ \citenamefont {Klein}}]{Juz05PRA}%
  \BibitemOpen
  \bibfield  {author} {\bibinfo {author} {\bibfnamefont {G.}~\bibnamefont {Juzeli\ifmmode~\bar{u}\else \={u}\fi{}nas}}, \bibinfo {author} {\bibfnamefont {P.}~\bibnamefont {\"Ohberg}}, \bibinfo {author} {\bibfnamefont {J.}~\bibnamefont {Ruseckas}},\ and\ \bibinfo {author} {\bibfnamefont {A.}~\bibnamefont {Klein}},\ }\bibfield  {title} {\bibinfo {title} {Effective magnetic fields in degenerate atomic gases induced by light beams with orbital angular momenta},\ }\href {https://doi.org/10.1103/PhysRevA.71.053614} {\bibfield  {journal} {\bibinfo  {journal} {Phys. Rev. A}\ }\textbf {\bibinfo {volume} {71}},\ \bibinfo {pages} {053614} (\bibinfo {year} {2005})}\BibitemShut {NoStop}%
\bibitem [{\citenamefont {Juzeli{\={u}}nas}\ \emph {et~al.}(2005)\citenamefont {Juzeli{\={u}}nas}, \citenamefont {Ruseckas},\ and\ \citenamefont {{\"O}hberg}}]{Juz05JPB}%
  \BibitemOpen
  \bibfield  {author} {\bibinfo {author} {\bibfnamefont {G.}~\bibnamefont {Juzeli{\={u}}nas}}, \bibinfo {author} {\bibfnamefont {J.}~\bibnamefont {Ruseckas}},\ and\ \bibinfo {author} {\bibfnamefont {P.}~\bibnamefont {{\"O}hberg}},\ }\bibfield  {title} {\bibinfo {title} {Effective magnetic fields induced by {EIT} in ultra-cold atomic gases},\ }\href {https://doi.org/10.1088/0953-4075/38/23/001} {\bibfield  {journal} {\bibinfo  {journal} {Journal of Physics B: Atomic, Molecular and Optical Physics}\ }\textbf {\bibinfo {volume} {38}},\ \bibinfo {pages} {4171} (\bibinfo {year} {2005})}\BibitemShut {NoStop}%
\bibitem [{\citenamefont {Juzeli\=unas}\ and\ \citenamefont {Spielman}(2012)}]{Juz-Spielm2012NJP}%
  \BibitemOpen
  \bibfield  {author} {\bibinfo {author} {\bibfnamefont {G.}~\bibnamefont {Juzeli\=unas}}\ and\ \bibinfo {author} {\bibfnamefont {I.~B.}\ \bibnamefont {Spielman}},\ }\bibfield  {title} {\bibinfo {title} {Flux lattices reformulated},\ }\href {https://doi.org/10.1088/1367-2630/14/12/123022} {\bibfield  {journal} {\bibinfo  {journal} {New. J. Phys.}\ }\textbf {\bibinfo {volume} {14}},\ \bibinfo {pages} {123022} (\bibinfo {year} {2012})}\BibitemShut {NoStop}%
\bibitem [{\citenamefont {Cooper}(2011)}]{Cooper11PRL}%
  \BibitemOpen
  \bibfield  {author} {\bibinfo {author} {\bibfnamefont {N.~R.}\ \bibnamefont {Cooper}},\ }\bibfield  {title} {\bibinfo {title} {Optical flux lattices for ultracold atomic gases},\ }\href {https://doi.org/10.1103/PhysRevLett.106.175301} {\bibfield  {journal} {\bibinfo  {journal} {Phys. Rev. Lett.}\ }\textbf {\bibinfo {volume} {106}},\ \bibinfo {pages} {175301} (\bibinfo {year} {2011})}\BibitemShut {NoStop}%
\bibitem [{\citenamefont {Cooper}\ and\ \citenamefont {Dalibard}(2013)}]{Cooper-Dalibard13PRL}%
  \BibitemOpen
  \bibfield  {author} {\bibinfo {author} {\bibfnamefont {N.~R.}\ \bibnamefont {Cooper}}\ and\ \bibinfo {author} {\bibfnamefont {J.}~\bibnamefont {Dalibard}},\ }\bibfield  {title} {\bibinfo {title} {Reaching fractional quantum {H}all states with optical flux lattices},\ }\href {https://doi.org/10.1103/PhysRevLett.110.185301} {\bibfield  {journal} {\bibinfo  {journal} {Phys. Rev. Lett.}\ }\textbf {\bibinfo {volume} {110}},\ \bibinfo {pages} {185301} (\bibinfo {year} {2013})}\BibitemShut {NoStop}%
\bibitem [{\citenamefont {Haldane}(1988)}]{Haldane88PRL}%
  \BibitemOpen
  \bibfield  {author} {\bibinfo {author} {\bibfnamefont {F.~D.~M.}\ \bibnamefont {Haldane}},\ }\bibfield  {title} {\bibinfo {title} {Model for a quantum {H}all effect without {L}andau levels: {C}ondensed-matter realization of the "parity anomaly"},\ }\href {https://doi.org/10.1103/PhysRevLett.61.2015} {\bibfield  {journal} {\bibinfo  {journal} {Phys. Rev. Lett.}\ }\textbf {\bibinfo {volume} {61}},\ \bibinfo {pages} {2015} (\bibinfo {year} {1988})}\BibitemShut {NoStop}%
\bibitem [{\citenamefont {Chiu}\ \emph {et~al.}(2016)\citenamefont {Chiu}, \citenamefont {Teo}, \citenamefont {Schnyder},\ and\ \citenamefont {Ryu}}]{Chiu16RMP}%
  \BibitemOpen
  \bibfield  {author} {\bibinfo {author} {\bibfnamefont {C.-K.}\ \bibnamefont {Chiu}}, \bibinfo {author} {\bibfnamefont {J.~C.~Y.}\ \bibnamefont {Teo}}, \bibinfo {author} {\bibfnamefont {A.~P.}\ \bibnamefont {Schnyder}},\ and\ \bibinfo {author} {\bibfnamefont {S.}~\bibnamefont {Ryu}},\ }\bibfield  {title} {\bibinfo {title} {Classification of topological quantum matter with symmetries},\ }\href {https://doi.org/10.1103/RevModPhys.88.035005} {\bibfield  {journal} {\bibinfo  {journal} {Rev. Mod. Phys.}\ }\textbf {\bibinfo {volume} {88}},\ \bibinfo {pages} {035005} (\bibinfo {year} {2016})}\BibitemShut {NoStop}%
\bibitem [{\citenamefont {Braver}\ \emph {et~al.}(2025)\citenamefont {Braver}, \citenamefont {Burba}, \citenamefont {Nair}, \citenamefont {{\v Z}labys}, \citenamefont {Anisimovas}, \citenamefont {Busch},\ and\ \citenamefont {Juzeli{\=u}nas}}]{Braver2025}%
  \BibitemOpen
  \bibfield  {author} {\bibinfo {author} {\bibfnamefont {Y.}~\bibnamefont {Braver}}, \bibinfo {author} {\bibfnamefont {D.}~\bibnamefont {Burba}}, \bibinfo {author} {\bibfnamefont {S.~S.}\ \bibnamefont {Nair}}, \bibinfo {author} {\bibfnamefont {G.}~\bibnamefont {{\v Z}labys}}, \bibinfo {author} {\bibfnamefont {E.}~\bibnamefont {Anisimovas}}, \bibinfo {author} {\bibfnamefont {T.}~\bibnamefont {Busch}},\ and\ \bibinfo {author} {\bibfnamefont {G.}~\bibnamefont {Juzeli{\=u}nas}},\ }\href {https://doi.org/10.48550/arXiv.2506.08683} {\bibinfo {title} {Light-induced localized vortices in multicomponent {B}ose-{E}instein condensates}} (\bibinfo {year} {2025}),\ \Eprint {https://arxiv.org/abs/2506.08683} {arXiv:2506.08683 [cond-mat.quant-gas]} \BibitemShut {NoStop}%
\bibitem [{Note2()}]{Note2}%
  \BibitemOpen
  \bibinfo {note} {We have calculated the Chern numbers $c$ of the Bloch bands using the algorithm presented by T. Fukui, Y. Hatsugai, and H. Suzuki, Chern Numbers in Discretized Brillouin Zone: Efficient Method of Computing (Spin) Hall Conductances, J. Phys. Soc. Jpn. \protect \textbf {74}, 1674 (2005).}\BibitemShut {Stop}%
\bibitem [{\citenamefont {L{\'e}onard}\ \emph {et~al.}(2023)\citenamefont {L{\'e}onard}, \citenamefont {Kim}, \citenamefont {Kwan}, \citenamefont {Segura}, \citenamefont {Grusdt}, \citenamefont {Repellin}, \citenamefont {Goldman},\ and\ \citenamefont {Greiner}}]{Greiner23Nature}%
  \BibitemOpen
  \bibfield  {author} {\bibinfo {author} {\bibfnamefont {J.}~\bibnamefont {L{\'e}onard}}, \bibinfo {author} {\bibfnamefont {S.}~\bibnamefont {Kim}}, \bibinfo {author} {\bibfnamefont {J.}~\bibnamefont {Kwan}}, \bibinfo {author} {\bibfnamefont {P.}~\bibnamefont {Segura}}, \bibinfo {author} {\bibfnamefont {F.}~\bibnamefont {Grusdt}}, \bibinfo {author} {\bibfnamefont {C.}~\bibnamefont {Repellin}}, \bibinfo {author} {\bibfnamefont {N.}~\bibnamefont {Goldman}},\ and\ \bibinfo {author} {\bibfnamefont {M.}~\bibnamefont {Greiner}},\ }\bibfield  {title} {\bibinfo {title} {Realization of a fractional quantum {H}all state with ultracold atoms},\ }\href {https://doi.org/10.1038/s41586-023-06122-4} {\bibfield  {journal} {\bibinfo  {journal} {Nature}\ }\textbf {\bibinfo {volume} {619}},\ \bibinfo {pages} {495} (\bibinfo {year} {2023})}\BibitemShut {NoStop}%
\bibitem [{\citenamefont {Lunt}\ \emph {et~al.}(2024)\citenamefont {Lunt}, \citenamefont {Hill}, \citenamefont {Reiter}, \citenamefont {Preiss}, \citenamefont {Ga{\l}ka},\ and\ \citenamefont {Jochim}}]{Joachim24PRL}%
  \BibitemOpen
  \bibfield  {author} {\bibinfo {author} {\bibfnamefont {P.}~\bibnamefont {Lunt}}, \bibinfo {author} {\bibfnamefont {P.}~\bibnamefont {Hill}}, \bibinfo {author} {\bibfnamefont {J.}~\bibnamefont {Reiter}}, \bibinfo {author} {\bibfnamefont {P.~M.}\ \bibnamefont {Preiss}}, \bibinfo {author} {\bibfnamefont {M.}~\bibnamefont {Ga{\l}ka}},\ and\ \bibinfo {author} {\bibfnamefont {S.}~\bibnamefont {Jochim}},\ }\bibfield  {title} {\bibinfo {title} {Realization of a {L}aughlin state of two rapidly rotating fermions},\ }\href {https://doi.org/10.1103/PhysRevLett.133.253401} {\bibfield  {journal} {\bibinfo  {journal} {Physical Review Letters}\ }\textbf {\bibinfo {volume} {133}},\ \bibinfo {pages} {253401} (\bibinfo {year} {2024})}\BibitemShut {NoStop}%
\bibitem [{\citenamefont {Wang}\ \emph {et~al.}(2021)\citenamefont {Wang}, \citenamefont {Cano}, \citenamefont {Millis}, \citenamefont {Liu},\ and\ \citenamefont {Yang}}]{Wang21PRL}%
  \BibitemOpen
  \bibfield  {author} {\bibinfo {author} {\bibfnamefont {J.}~\bibnamefont {Wang}}, \bibinfo {author} {\bibfnamefont {J.}~\bibnamefont {Cano}}, \bibinfo {author} {\bibfnamefont {A.~J.}\ \bibnamefont {Millis}}, \bibinfo {author} {\bibfnamefont {Z.}~\bibnamefont {Liu}},\ and\ \bibinfo {author} {\bibfnamefont {B.}~\bibnamefont {Yang}},\ }\bibfield  {title} {\bibinfo {title} {Exact {L}andau level description of geometry and interaction in a flatband},\ }\href {https://doi.org/10.1103/PhysRevLett.127.246403} {\bibfield  {journal} {\bibinfo  {journal} {Phys. Rev. Lett.}\ }\textbf {\bibinfo {volume} {127}},\ \bibinfo {pages} {246403} (\bibinfo {year} {2021})}\BibitemShut {NoStop}%
\bibitem [{\citenamefont {Estienne}\ \emph {et~al.}(2023)\citenamefont {Estienne}, \citenamefont {Regnault},\ and\ \citenamefont {Cr\'epel}}]{Crepel2023PRR}%
  \BibitemOpen
  \bibfield  {author} {\bibinfo {author} {\bibfnamefont {B.}~\bibnamefont {Estienne}}, \bibinfo {author} {\bibfnamefont {N.}~\bibnamefont {Regnault}},\ and\ \bibinfo {author} {\bibfnamefont {V.}~\bibnamefont {Cr\'epel}},\ }\bibfield  {title} {\bibinfo {title} {Ideal {C}hern bands as {L}andau levels in curved space},\ }\href {https://doi.org/10.1103/PhysRevResearch.5.L032048} {\bibfield  {journal} {\bibinfo  {journal} {Phys. Rev. Res.}\ }\textbf {\bibinfo {volume} {5}},\ \bibinfo {pages} {L032048} (\bibinfo {year} {2023})}\BibitemShut {NoStop}%
\bibitem [{\citenamefont {Ledwith}\ \emph {et~al.}(2023)\citenamefont {Ledwith}, \citenamefont {Vishwanath},\ and\ \citenamefont {Parker}}]{Ledwith23PRB}%
  \BibitemOpen
  \bibfield  {author} {\bibinfo {author} {\bibfnamefont {P.~J.}\ \bibnamefont {Ledwith}}, \bibinfo {author} {\bibfnamefont {A.}~\bibnamefont {Vishwanath}},\ and\ \bibinfo {author} {\bibfnamefont {D.~E.}\ \bibnamefont {Parker}},\ }\bibfield  {title} {\bibinfo {title} {Vortexability: A unifying criterion for ideal fractional {C}hern insulators},\ }\href {https://doi.org/10.1103/PhysRevB.108.205144} {\bibfield  {journal} {\bibinfo  {journal} {Phys. Rev. B}\ }\textbf {\bibinfo {volume} {108}},\ \bibinfo {pages} {205144} (\bibinfo {year} {2023})}\BibitemShut {NoStop}%
\bibitem [{\citenamefont {Mitscherling}\ \emph {et~al.}(2025)\citenamefont {Mitscherling}, \citenamefont {Avdoshkin},\ and\ \citenamefont {Moore}}]{Mitscherling2025arXiv}%
  \BibitemOpen
  \bibfield  {author} {\bibinfo {author} {\bibfnamefont {J.}~\bibnamefont {Mitscherling}}, \bibinfo {author} {\bibfnamefont {A.}~\bibnamefont {Avdoshkin}},\ and\ \bibinfo {author} {\bibfnamefont {J.~E.}\ \bibnamefont {Moore}},\ }\href {https://arxiv.org/abs/2412.03637} {\bibinfo {title} {Gauge-invariant projector calculus for quantum state geometry and applications to observables in crystals}} (\bibinfo {year} {2025}),\ \Eprint {https://arxiv.org/abs/2412.03637} {arXiv:2412.03637 [cond-mat.str-el]} \BibitemShut {NoStop}%
\end{thebibliography}%

\end{document}